\documentclass[pss]{wiley2sp} 
\usepackage{amsmath}
\usepackage{bm}              
\usepackage{w-greek}
\usepackage{graphicx}         
\usepackage{hyperref}

\newcommand{\ket}[1]{\left|#1\right>}
\newcommand{\bra}[1]{\left<#1\right|}

\newcommand{\expval}[1]{\left< #1 \right>}
\newcommand{\nn}{\nonumber\\}

\newcommand{\f}[1]{\mbox{\boldmath$#1$}}

\newcommand{\bea}{\begin{eqnarray}}
\newcommand{\eea}{\end{eqnarray}}

\newcommand{\abs}[1]{{\left| #1 \right|}}
\newcommand{\trace}[1]{{\rm Tr}\left\{ #1 \right\}}

\newcommand{\ii}{{\rm i}}

\begin{document}

\title{Feedback-charging a metallic island}

\titlerunning{Feedback-charging a metallic island}

\author{Gernot Schaller\textsuperscript{\Ast,\textsf{\bfseries 1}}}

\authorrunning{G. Schaller}

\mail{e-mail
  \textsf{gernot.schaller@tu-berlin.de}, Phone:
  +49-30-31421777, Fax: +49-30-31421130}

\institute{\textsuperscript{1}\,Institut f\"ur Theoretische Physik, Technische Universit\"at Berlin, Hardenbergstr. 36, D-10623 Berlin, Germany}

\received{XXXX, revised XXXX, accepted XXXX} 
\published{XXXX} 


\abstract{We consider electronic transport through a single-electron quantum dot that is tunnel-coupled to 
an electronic lead and a metallic island.
A background reservoir keeps the metallic island at a thermal state with the ambient temperature, while
the charge accumulated on the island is reflected in a time-dependent chemical potential.
Without feedback, a current would flow through the system until the chemical potentials of island and
lead are equilibrated.

A feedback loop can be implemented by a quantum point contact detecting the dot state, 
classical processing of the result and appropriate feedback actions on the electronic tunneling rates 
taken, with the objective to direct the current in a preferred direction.
Since we directly take the detector counting statistics into account, this automatically includes measurement
errors in the description.
When mainly the rates are modified but hardly any energy is exchanged with the system, this feedback
loop effectively implements a Maxwell demon, capable of transporting electrons against an electric bias
and thereby charging the metallic island.
Once the feedback protocol is stopped, the metallic island simply discharges.

We find that a quantitative detector model may be useful for a realistic statistical description of feedback loops.
}

\titlefigure[width=0.45\textwidth,clip=true]{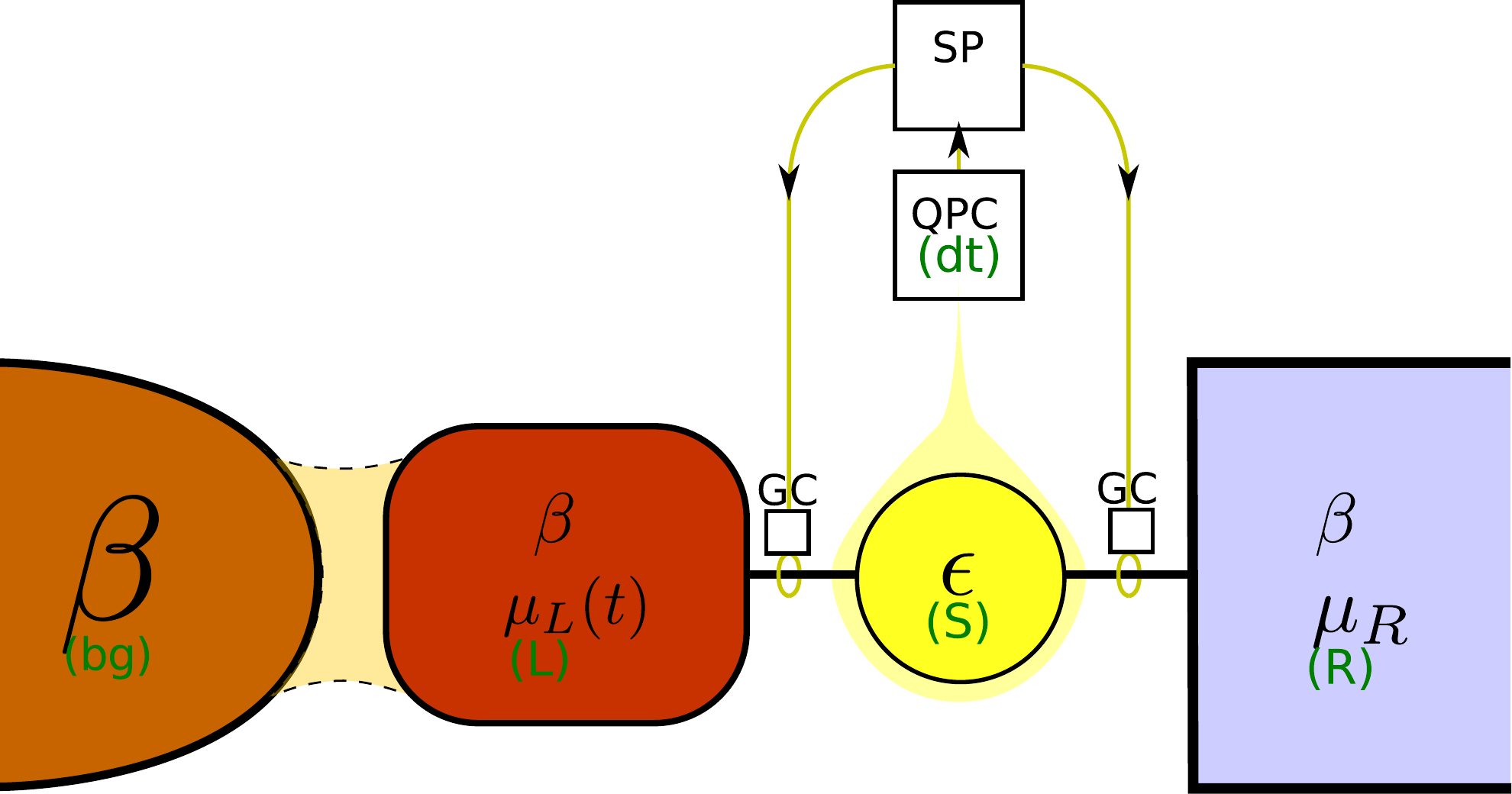}
\titlefigurecaption{\label{FIG:sketch_model}
Sketch of the model. 
The system (S) consists of a single-electron transistor (SET) dot that can be occupied with at most a single electron.
It is tunnel-coupled to a lead (R) held at a fixed thermal equilibrium state and to a metallic
island (L).
The metallic island is held at a fixed temperature by an additional background reservoir (bg), with which it can exchange energy but
no particles (shaded contact region). 
Its chemical potential would normally evolve until $\mu_L(t) \to \mu_R$.
A simple feedback loop can be implemented measuring the state of the system with a quantum point contact (dt), 
classical signal processing (SP), and using the gate controls (GC) to obtain occupation-dependent tunneling rates.
With feedback active, it is possible to charge the island ($\mu_L \neq \mu_R$).
Expressions in brackets (green) denote abbreviations used in the paper.
}

\maketitle

\section{Introduction}

The concept of feedback control has been used for a long time to automate processes.
A prominent example is the centrifugal governor used in steam engines to regulate their speed.
Nowadays, feedback circuits are part of many appliances~\cite{astrom2008} such as e.g. in the automatic speed control of cars, 
heating control, and automatic mowers.

In the control of quantum systems, feedback loops can be used to enhance desired properties, e.g. to help fight decoherence or to 
create desired states~\cite{korotkov2001a,wiseman2010,kiesslich2011a,sayrin2011a}.
Also in electronic transport, feedback has been proposed~\cite{brandes2010a} and used~\cite{wagner2016a} to control
the width of otherwise dispersing probability distributions.
Due to the many different areas of application, feedback comes in many variants.
However, all of them include that in some way information obtained from the system is fed back into the system to 
obtain some optimized behaviour.
For quantum systems, one way of realizing an external feedback loop is by a (weak or strong) measurement, classical processing, and subsequent
control actions performed on the system.
Alternatively, one can attach another quantum system to it, which by the physical interaction between the two ideally forces
the original quantum system into the desired behaviour, which realizes an autonomous feedback loop.

A specific external feedback loop that goes back to the foundations of thermodynamics is the famous Maxwell demon~\cite{leff1990,maruyama2009a}.
It describes an intelligent being capable of monitoring the speed of molecules inside a two-component box.
Depending on the speed of individual molecules, the demon opens or closes a shutter and thereby allows or denies the molecule
to transfer from one compartment of the box to the other. 
When operated for sufficiently long time, the demon can sort the molecules into a fast and slow fraction and thereby generates a thermal gradient.
While ideally leaving the energy balance unaffected (if opening or closing the shutter does not cost energy), it will locally reduce the 
entropy of the system.
Eventually, the created thermal gradient may be used to generate work from information.
The apparent violation of the second law of thermodynamics can be globally resolved by including the entropy balance of the demon, 
which -- to remain operable -- needs to delete information and thereby generates entropy exceeding the local reduction in the system~\cite{landauer1961a}.
However, also from the local perspective of the system a modified second law may be established, taking an effective
information current entering the system into account.

In recent years, techniques for the measurement and manipulation of single-electron devices have been continuously improved.
In particular for electronic transport systems, it is now possible to measure the transport of single charges with extreme 
precision~\cite{gustavsson2006a,sukhorukov2007a}.
This has inspired the investigation of fundamental symmetries related to the second law such as fluctuation theorems~\cite{andrieux2006a,esposito2007a,harbola2007a,golubev2011a}, 
also from the experimental perspective~\cite{saira2012a}.
These fundamental thermodynamic relations are modified in presence of feedback~\cite{sagawa2008a,horowitz2010a,schaller2011b,lahiri2012a,sagawa2012b}.
Therefore, electronic quantum dot systems have been suggested as a very promising test-bed to approach the Maxwell demon scenario~\cite{averin2011a,schaller2011b}.
Even autonomous versions of a Maxwell demon feedback loop have been suggested~\cite{strasberg2013a} and implemented~\cite{koski2015a}.

If at all, most works on external feedback loops consider however the measurement errors by the demon only in an effective way. 
Here, we therefore aim to include the detector in the description using a very simple model.
Furthermore, for small nano-devices many reservoirs are actually of finite size, such that they cannot be regarded as 
fully stationary.
This means for example that they may be affected by the presence of a demon building up charge in them.
The present paper attempts to provide some steps to fill this gap.

It is organized as follows: 
We first introduce the model in Sec.~\ref{SEC:model} starting from a microscopic discussion in Sec.~\ref{SEC:hamiltonian} and then state the
resulting generalized rate equation describing the counting statistics of system and detector charges in Sec.~\ref{SEC:master}.
Then, we consider piecewise-constant driving in absence of measurement errors in Sec.~\ref{SEC:piecewise_constant}.
Here, we discuss how the feedback changes the energetic and entropic balance of the system in Sec.~\ref{SEC:entropic_balance} and how the
matter current leaving the system to the metallic island affects its chemical potential in Sec.~\ref{SEC:islandcharging}. 
We present our results in Sec.~\ref{SEC:results}, starting with a trajectory-based implementation of the feedback loop in Sec.~\ref{SEC:feedback}.
We then discuss how the feedback loop can be described on average, leading to evolution equations for the system in Sec.~\ref{SEC:average_system} and
for the island potential in Sec.~\ref{SEC:average_potential}.
Finally, we discuss some experimental difficulties in Sec.~\ref{SEC:experiments}, further operational modes for the considered model in Sec.~\ref{SEC:applications},  
and conclude with a summary in Sec.~\ref{SEC:summary}.

Furthermore, we provide technical tools such as counting statistics in App.~\ref{APP:fcs}, error probabilities for the 
charge detector in App.~\ref{APP:errors} and relate weak and strong measurement with the counting statistics in App.~\ref{APP:measurements}.
We also also address fundamentals as canonical vs. grand-canonical treatment in App.~\ref{APP:canonical}
and generic entropic balances for piecewise-constant feedbacks in App.~\ref{APP:entropy_general}.

\section{Model}\label{SEC:model}

\subsection{Hamiltonian}\label{SEC:hamiltonian}

The full Hamiltonian of the model (compare also the figure on the title page) can be additively decomposed into contributions 
from the single-electron transistor (SET) system (S), a right reservoir (R), a metallic island (L), a background reservoir (bg), a detector (dt), 
tunneling-type interactions between system and right reservoir (SR) and system and metallic island (SL), and
finally interactions (excluding particle tunneling) between the system and the detector (S,dt), 
and the background reservoir and the metallic island (bg,L), respectively, 
\bea\label{EQ:hams}
H &=& H_S + H_R + H_L + H_{\rm bg} + H_{\rm dt}\nn
&& +H_{SR} + H_{SL} + H_{S, \rm dt} + H_{\rm bg,L}\,.
\eea
The first individual contributions are
\bea
H_S &=& \epsilon d^\dagger d\,,\nn
H_R &=& \sum_{k} \epsilon_{kR} c_{kR}^\dagger c_{kR}\,,\qquad
H_L = \sum_{k} \epsilon_{kL} c_{kL}^\dagger c_{kL}\,,
\eea
where $\epsilon$ denotes the dot level, $\epsilon_{k\alpha}$ the energy of mode $k$ in the right lead ($\alpha=R$)
or the metallic island ($\alpha=L$), and occupied states of these modes are created with the fermionic operators $d^\dagger$, 
$c_{kR}^\dagger$, and $c_{kL}^\dagger$, respectively.
We note that spin degrees of freedom are neglected.
The quantum dot is tunnel-coupled to the right lead ($\alpha=R$) and the left metallic island ($\alpha=L$)
\bea
H_{S\alpha} &=& \sum_{k} \left(t_{k\alpha} d c_{k\alpha}^\dagger + {\rm h.c.}\right)
\eea
with tunneling amplitude $t_{k\alpha}$.
Adopting a continuum representation in the lead and the metallic island, treating their tunnel-interactions perturbatively, and
for the moment assuming the island potential to be constant, 
one obtains for the steady-state current through the system (counting positive when directed from left to right)~\cite{esposito2009b}
\bea\label{EQ:sscur_set}
\bar{I}_M = \frac{\Gamma_L \Gamma_R}{\Gamma_L+\Gamma_R} (f_L-f_R)\,,
\eea
where $\Gamma_\alpha=\Gamma_\alpha(\epsilon)$ with 
\mbox{$\Gamma_\alpha(\omega)=2\pi\sum_k \abs{t_{k\alpha}}^2 \allowbreak \delta(\omega-\epsilon_{k\alpha})$}
denote tunneling rates and 
$f_\alpha=[e^{\beta(\epsilon-\mu_\alpha)}+1]^{-1}$ the Fermi functions of the right lead (R) or the metallic 
island (L) evaluated at the dot energy.

Later-on, we will allow for simple time-dependencies of the tunneling rates $\Gamma_\alpha$ and system parameters $\epsilon$ 
due to feedback measures taken and a time-dependent island potential $\mu_L$ due to charging effects.

The actual form of the background Hamiltonian and its interaction with the metallic island is quite arbitrary.
However, we
demand that the background reservoirs can exchange energy with the metallic island ($[H_{\rm bg,L}, H_L]\neq 0$) but
no particles $[H_{\rm bg,L}, N_L]=0$ with $N_L=\sum_k c_{kL}^\dagger c_{kL}$ being the particle number operator of the island.
This would be satisfied e.g. by electron-phonon scattering processes.
Under the usual assumptions applied in the derivation of master equations, this would drag the metallic island toward
a thermal state, characterized by the inverse temperature $\beta$ of the background reservoir and a chemical potential that
is determined by the other junctions of the metallic island.
When we demand in addition that the background reservoir thermalizes the metallic island on a timescale that is much shorter
than the dynamics of the system, we can approximate the state
of the metallic island by a time-dependent thermal state~\cite{schaller2014b}
in the grand-canonical ensemble
\bea\label{EQ:thermalstate}
\rho_L(t) \propto e^{-\beta (H_{L} - \mu_L(t) N_{L})}\,.
\eea
Here -- owing to the charge transport -- the chemical potential of the metallic island is time-dependent.
The actual value of the potential will have to be determined self-consistently from the current entering the island.
Such a treatment requires that the metallic island is small enough such that is can be influenced by the charge transfer through the dot in finite
time, but at the same time also large enough such that a continuum description applies.
On the other hand, in the advertised limit it would also be applicable to treat the metallic island using different canonical ensembles
for different fixed particle numbers.
We discuss this possibility in Appendix~\ref{APP:canonical}, where we show that for a continuum of frequencies in the island the difference between
the two approaches is negligible.

Charge detection in quantum dots is typically implemented by a quantum point contact~\cite{vanhouten1996a}, and a simple model
for such a device is given by two leads that are directly tunnel-coupled~\cite{levinson1997a,elattari2000a,golubev2011a}
\bea
H_{\rm dt} &=& \sum_k \varepsilon_{k1} d_{k1}^\dagger d_{k1} + \sum_k \varepsilon_{k2} d_{k2}^\dagger d_{k2}\nn
&&+ \sum_{kk'} \left[\tau_{kk'} d_{k1} d_{k'2}^\dagger + {\rm h.c.}\right]
\eea
with tunneling amplitude $\tau_{kk'}$ describing an electron transfer from mode $k$ of lead $1$ into mode $k'$ of lead $2$.
This tunneling process is modified when a charge is present in the SET system
\bea
H_{\rm S, dt} = d^\dagger d \sum_{kk'} \left[\Delta\tau_{kk'} d_{k1} d_{k'2}^\dagger + {\rm h.c.}\right]\,, 
\eea
where $\Delta \tau_{kk'}$ leads to a modification (usually a suppression) of the effective quantum point contact (QPC) current when the SET is occupied.

\subsection{Reduced dynamics}\label{SEC:master}

The dynamics of the quantum dot alone is well described by its time-dependent occupation (superpositions between 
electronic states on the dot and the other components are beyond the weak-coupling regime treated in this paper).
Beyond this, we are interested in the statistics of transferred charges to the left metallic island, 
the total number of dot configuration changes, and
the total number of charges transferred through the QPC circuit.
These questions can be addressed with tools from Full Counting Statistics~\cite{levitov1996a,bagrets2003a}, see also appendix~\ref{APP:fcs}.
Treating the tunneling $t_{k\alpha}$ between system and lead and between system and metallic island perturbatively and also
considering the QPC tunneling $\tau_{kk'}$ as well as $\Delta \tau_{kk'}$ (low transparency QPC) as weak, the 
desired quantities can be captured by the generalized rate equation (compare e.g. Ref.~\cite{schaller2014} for 
a similar setup)
\bea\label{EQ:liouvilliantotal}
{\cal L}(\f{\chi}) &=& {\cal L}_L(\chi,\lambda) + {\cal L}_R(\chi) + {\cal L}_{\rm dt}(\xi)\,,\nn
{\cal L}_L(\chi,\lambda) &=& \left(\begin{array}{cc}
-\Gamma_L f_L & +\Gamma_L (1-f_L) e^{\ii(\chi-\lambda)}\\
+\Gamma_L f_L e^{\ii(\chi+\lambda)} & -\Gamma_L (1-f_L)
\end{array}\right)\,,\nn
{\cal L}_R(\chi) &=& \left(\begin{array}{cc}
-\Gamma_R f_R & +\Gamma_R (1-f_R) e^{+\ii \chi}\\
+\Gamma_R f_R e^{+\ii\chi} & -\Gamma_R (1-f_R)
\end{array}\right)\,,\nn
{\cal L}_{\rm dt}(\xi) &=& \left(\gamma_f (e^{+\ii\xi}-1)+\gamma_b(e^{-\ii\xi}-1) \right)\left(\begin{array}{cc}
1 & 0\\
0 & \kappa 
\end{array}\right)\,,
\eea
where $\f{\chi}=(\chi,\lambda,\xi)$ and the counting fields $\chi$, $\lambda$, and $\xi$ can be used to infer the number of
total dot configuration changes, the net charges transferred between dot and metallic island, and the 
net charges transferred through the QPC, respectively, compare App.~\ref{APP:fcs} and App.~\ref{APP:errors}.
In the following, we will also use the notation ${\cal L}^{ij} = \left({\cal L}(\f{0})\right)_{ij}$ to denote the matrix elements
of the Liouvillian.
The rates $\gamma_{f/b}$ denote the forward/backward transmission rates for the QPC.
They are related to the microscopic parameters via~\cite{schaller2014}
\bea\label{EQ:qpc_parameters}
\gamma_f = \frac{T_0 V}{1-e^{-\beta V}}\,,\qquad
\gamma_b = \frac{T_0 V}{e^{+\beta V}-1}
\eea
with the QPC bias voltage $V$ and $T(\omega,\omega')=2\pi \times\allowbreak\sum_{kk'} \abs{\tau_{kk'}}^2\delta(\omega-\varepsilon_{k1})\delta(\omega'-\varepsilon_{k2}) \to T_0$
denoting the dimensionless transmission of the QPC in the wideband limit.
Finally, the dimensionless parameter $\kappa$ effectively describes the modification of the QPC current when the SET is occupied, which is microscopically implemented via
the relation $2\pi \sum_{kk'} \abs{\tau_{kk'}+\Delta \tau_{kk'}}^2\times\allowbreak \delta(\omega-\varepsilon_{k1})\delta(\omega'-\varepsilon_{k2}) \to \kappa T_0$.
In this paper, we will for simplicity only consider uni-directional QPC transport by assuming a sufficiently large bias voltage $\beta V \gg 1$, such that
$\gamma_b \to 0$ and also consider only a reduction of the QPC current $0<\kappa<1$.

To infer the dynamics of the SET population only without making reference to the counting statistics, 
it suffices to consider ${\cal L}(\f{0})$, where we see that in this limit, the QPC has to this order 
no effect as ${\cal L}_{\rm dt}(0)=\f{0}$.

\section{Piecewise-constant error-free feedback}\label{SEC:piecewise_constant}

Strictly speaking, the previously discussed rate equations~(\ref{EQ:liouvilliantotal}) hold only for time-independent systems.
In contrast, feedback requires time-dependent operations modifying the system energies and/or the tunneling rates.
Furthermore, the system currents may influence the island potential by charging effects.
Therefore, we consider here a specific simple form of piecewise-constant control and a way to compute the 
time-dependent chemical potential assuming an ultrafast relaxation to thermal equilibrium~(\ref{EQ:thermalstate}).

Then, when changes are assumed to happen instantaneously, the previously discussed rate equations are still
meaningful to describe the evolution right after a measurement.
Formally, in such a scenario, the average evolution under feedback ${\cal L}_{\rm fb}$ can be constructed from the conditional 
evolutions ${\cal L}_m$ via ${\cal L}_{\rm fb} = \sum_m {\cal L}_m {\cal P}_m$, 
where ${\cal P}_m$ correspond to the two measurement outcomes of measuring a high or low current, respectively.
We note that the driving need not be conditioned on the actual SET occupation but rather the measured one and may thus be erroneous.
Before considering the impact of errors in Sec.~\ref{SEC:results}, we first consider the impact on error-free feedback on
the thermodynamics in this section.

For systems with an external feedback loop there are many sources of entropy production that are not fully controllable.
Both measurement and control actions will be -- since executed by physically realistic devices -- associated with intrinsic contributions to both
the first and the second law.
Since we have introduced a QPC model for the charge detection in this paper, we can say a few words on the associated entropy production.
For such a detector, its average entropy production rate at QPC bias voltage $V$ and average QPC current
$I$ is given by $\dot{S}_{\ii}^{\rm dt} = \beta V I$.
In particular, since one requires a certain accuracy to make the feedback effective, a significant number of electrons needs to pass the 
measurement circuit, which leads to significant heat dissipation and associated entropy production already in this first step~\cite{barato2015a}.
Next, depending on the implementation, heat orders of magnitude larger will be produced in the signal processing step.
Finally, during the control step, heat may also be dissipated in the environment.
Quantifying all these effects would require a fully inclusive (i.e., autonomous) feedback control model~\cite{strasberg2013a,koski2015a}.

However, for our model, we can quantify the net effect the feedback has on the local entropy balance, which we will discuss below.

\subsection{Local entropic balance}\label{SEC:entropic_balance}

For error-free feedback~\cite{schaller2011b,esposito2012a},  
the average feedback rate matrix becomes (for simplicity without counting fields)
\bea\label{EQ:lmatfb}
{\cal L}_{\rm fb} = \sum_\alpha \left(\begin{array}{cc}
-\Gamma_\alpha^H f_\alpha^H & +\Gamma_\alpha^L [1-f_\alpha^L]\\
+\Gamma_\alpha^H f_\alpha^H & -\Gamma_\alpha^L [1-f_\alpha^L]
\end{array}\right)\,.
\eea
Here, the piecewise-constant driving leads to two possible values of the SET tunneling rates $\Gamma_\alpha\to \Gamma_\alpha^{H/L}$
and also of the system Hamiltonian ($\epsilon\to \epsilon^{H/L}$).
Since the dot parameters in the description only enter implicitly, we described the latter by conditional Fermi functions $f_\alpha \to f_\alpha^{H/L}$.
With such a feedback scheme, one will in general inject both energy and information into the system, which can be consistently 
treated on the local level.
In appendix~\ref{APP:entropy_general} we provide a general thermodynamic discussion of feedback-driven rate equations of the form $\dot{P}_i = \sum_j W_{ij}^{(j,\alpha)} P_j$, 
whereas in this section we analyze the specifics of our model.

Assuming the conditioned dot Hamiltonian as $H_S = \epsilon_{H/L} d^\dagger d$,
the empty dot has energies $E_0^{(0)} = 0$ and $E_1^{(0)}=\epsilon_H$, and when filled, 
the system has energies $E_0^{(1)} = 0$ and $E_1^{(1)} = \epsilon_L$.
Therefore, we can identify the heat entering the system from reservoir $\alpha$ during a jump out of the system as
$\Delta Q_{\rm out}^{(\alpha)} = E_0^{(1)}-E_1^{(1)}-\mu_\alpha (N_0-N_1) = -\epsilon_L+\mu$ and for a jump into the system 
as $\Delta Q_{\rm in}^{(\alpha)} = E_1^{(0)}-E_0^{(0)}-\mu_\alpha (N_1-N_0) = +\epsilon_H-\mu$, leading to an overall heat current of
\bea\label{EQ:heat_specific}
\dot{Q}^{(\alpha)} &=& -(\epsilon_L-\mu_\alpha) {\cal L}_{\rm fb}^{01,\alpha} P_1 + (\epsilon_H-\mu_\alpha) {\cal L}_{\rm fb}^{10,\alpha} P_0\nn
&=& I_E^{(\alpha)} - \mu_\alpha I_M^{(\alpha)}\,,
\eea
which also defines energy $I_E^{(\alpha)}$ and matter $I_M^{(\alpha)}$ currents entering the system from reservoir $\alpha$.
A similar result holds if also the energy of the empty state is changed by the feedback.
We can show (see App.\ref{APP:entropy_general}) that the energy change of the system is balanced by the energy 
currents entering the system from both reservoirs and the energy current injected by the feedback
\bea
I_E^{\rm fb} = (\epsilon_L-\epsilon_H) \sum_\alpha {\cal L}_{\rm fb}^{10,\alpha} P_0\,.
\eea

To discuss the entropic balance, we can with Eq.~(\ref{EQ:lmatfb}) write the ratio of backward- and forward rates for each reservoir as
\bea
\frac{{\cal L}_{\rm fb}^{01,\alpha}}{{\cal L}_{\rm fb}^{10,\alpha}} &=& \frac{\Gamma_\alpha^L}{\Gamma_\alpha^H} \frac{1-f_\alpha^L}{f_\alpha^H}
= \left(\frac{1-f_\alpha^H}{f_\alpha^H}\right) \left[\frac{\Gamma_\alpha^L}{\Gamma_\alpha^H}\right] \left\{\frac{1-f_\alpha^L}{1-f_\alpha^H}\right\}\,,\nn
\frac{{\cal L}_{\rm fb}^{10,\alpha}}{{\cal L}_{\rm fb}^{01,\alpha}}&=& \frac{\Gamma_\alpha^H}{\Gamma_\alpha^L} \frac{f_\alpha^H}{1-f_\alpha^L}
= \left(\frac{f_\alpha^L}{1-f_\alpha^L}\right) \left[\frac{\Gamma_\alpha^H}{\Gamma_\alpha^L}\right] \left\{\frac{f_\alpha^H}{f_\alpha^L}\right\}\,,
\eea
where we see from $(1-f_\alpha^H)/f_\alpha^H=e^{+\beta_\alpha(\epsilon_H-\mu_\alpha)}$ and $f_\alpha^L/(1-f_\alpha^L) = e^{-\beta_\alpha(\epsilon_L-\mu_\alpha)}$ that
the terms in round parentheses $(\ldots)$ will when inserted in the ``entropy flow'' term 
\bea
\dot{S}_{\rm e}^{(\alpha)} = \sum_{ij} W_{ij}^{(j,\alpha)} P_j \ln \frac{W_{ji}^{(i,\alpha)}}{W_{ij}^{(j,\alpha)}}
\eea
compose the entropy change in the reservoirs $-\beta_\alpha \dot{Q}^{(\alpha)}$, compare Eq.~(\ref{EQ:heat_specific}).
The terms in square brackets $[\ldots]$ are a pure Maxwell-demon contribution~\cite{esposito2012a} in the sense that they only affect the entropic balance directly, 
and the terms in curly brackets $\{\ldots\}$ describe the influence on the feedback energy injection on the entropic balance.
We therefore define the feedback parameters
\bea
\Delta_{01}^{(\alpha)} &=& \ln \frac{\Gamma_\alpha^L}{\Gamma_\alpha^H}\,,\qquad
\Delta_{10}^{(\alpha)} = \ln \frac{\Gamma_\alpha^H}{\Gamma_\alpha^L}\,,\nn
\sigma_{01}^{(\alpha)} &=& \ln \frac{f_\alpha^L}{f_\alpha^H}\,,\qquad
\sigma_{10}^{(\alpha)} = \ln \frac{1-f_\alpha^H}{1-f_\alpha^L}\,,
\eea
compare also Eq.~(\ref{EQ:detailed_balance_feedback}) in appendix~\ref{APP:entropy_general}.
We see that the information contribution of the feedback obeys $\Delta_{01}^{(\alpha)} = - \Delta_{10}^{(\alpha)}$ and
the energetic contribution obeys \mbox{$\sigma_{01}^{(\alpha)} \sigma_{10}^{(\alpha)} = \beta_\alpha (\epsilon_H-\epsilon_L)$}.
With these, the ``entropy flow'' term becomes modified by information currents $\dot{S}_{\rm e} = \sum_\alpha \beta_\alpha \dot{Q}^{(\alpha)} - {\cal I}_1 - {\cal I}_2$, 
which read explicitly
\bea
{\cal I}_1 &=& \sum_\alpha \left[{\cal L}_{\rm fb}^{01,\alpha} P_1 - {\cal L}_{\rm fb}^{10,\alpha} P_0\right] \ln \frac{\Gamma_\alpha^L}{\Gamma_\alpha^H}\,,\nn
{\cal I}_2 &=& \sum_\alpha \left[{\cal L}_{\rm fb}^{01,\alpha} P_1 \ln \frac{f_\alpha^L}{f_\alpha^H} + {\cal L}_{\rm fb}^{10,\alpha} P_0\ln \frac{1-f_\alpha^H}{1-f_\alpha^L}\right]\,.
\eea
Above, it is visible that the information current ${\cal I}_1$ is tightly coupled to the matter current.
At steady state, we have conservation of the matter currents $I_M = I_M^{(L)}=-I_M^{(R)}$ and from the first law also conservation of the
individual energy currents and the feedback energy current $I_E^{(L)}+I_E^{(R)} + I_E^{\rm fb}=0$.

Inserting these in the steady-state entropy production rate $\dot{S}_{\ii} = - \dot{S}_{\rm e}$ we find 
that at equal temperatures $\beta=\beta_L=\beta_R$ the second law reads
\bea\label{EQ:entropy_production_specific}
\dot{S}_{\ii} \to \beta(\mu_L-\mu_R) I_M +{\cal I}_1 + \beta I_E^{\rm fb} + {\cal I}_2 \ge 0\,.
\eea
Here, the first term contains the produced electric power $P=-(\mu_L-\mu_R) I_M$, which without feedback would always be negative.
The second term contains the purely informational contribution of the feedback to the entropic balance.
The third term quantifies how the difference of left and right energy currents $I_E^{(L)}+I_E^{(R)} =- I_E^{\rm fb}$ affects 
the heat exchanged with the reservoirs.
If the feedback does not affect the energy levels ($\epsilon_H=\epsilon_L$), this term will naturally vanish.
Finally, the last term describes the effect of the feedback level driving on the entropic balance.
Since the level driving also enters the entropic balance, we cannot interpret this simply as work on the system.

\begin{figure}[ht]
\includegraphics[width=0.48\textwidth,clip=true]{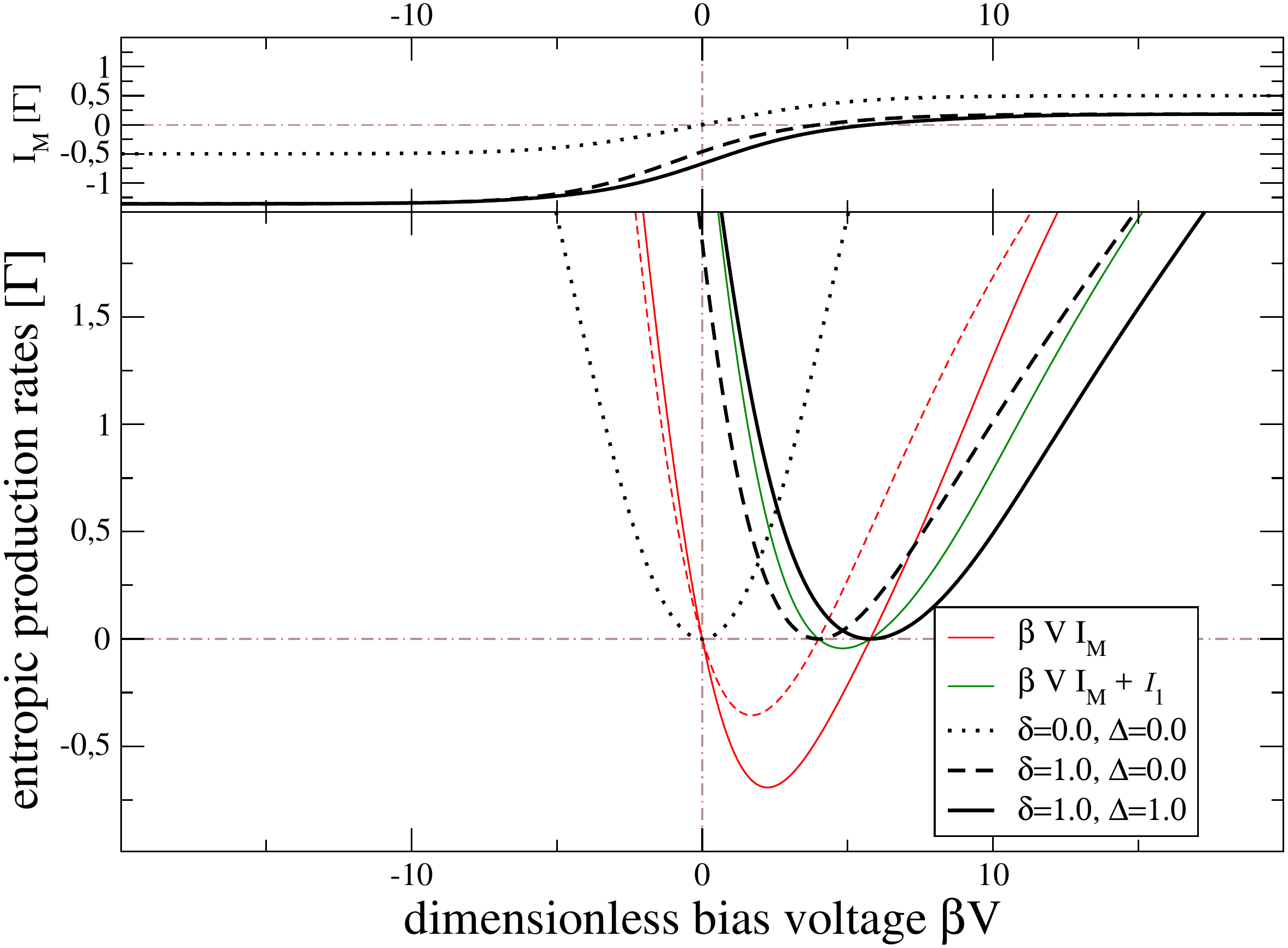}
\caption{\label{FIG:entropic_balance}
Plot of the matter current from left to right (top) and contributions to the total entropy production rate (\ref{EQ:entropy_production_specific})
(bottom) for situations without feedback $\delta=\Delta=0$ (dotted), 
with Maxwell-demon feedback $\delta=+1.0$, $\Delta=0$ (dashed), and with energy-injecting feedback $\delta=\Delta=+1.0$ (solid).
With feedback active (dashed and solid), we see that the matter current at equilibrium $V=0$ becomes negative and remains negative for a small region $0<V<V^*$, 
where the device produces positive power $P=-V I_M$ either 
using only information ($\Delta=0$) or information and energy injection ($\Delta\neq 0$).
Red thin curves of similar style denote the naive entropy production rate $\beta(\mu_L-\mu_R) I_M=-\beta P$ that one would conjecture in ignorance of any feedback actions taken.
Green thin curves of similar style denote the naive entropy production rate $-\beta(\mu_L-\mu_R) I_M + {\cal I}_1$ that one would conjecture when assuming that the feedback
does not affect the energy levels.
The black curves denote the true entropy production rate, which is positive in all parameter regimes.
Dash-dotted lines just serve for orientation. Other parameters: $\beta\epsilon=1$.
}
\end{figure}

For simplicity, we can parametrize the tunneling rates using only a single parameter
\bea
\Gamma_L^L &=& \Gamma e^{+\delta}\,,\qquad
\Gamma_R^L = \Gamma e^{-\delta}\,,\nn
\Gamma_L^H &=& \Gamma e^{-\delta}\,,\qquad
\Gamma_R^H = \Gamma e^{+\delta}\,,
\eea
which will for $\delta>0$ favor transport from right to left (preferentially charging the island).
This will not change the energetics, but the entropic balance is affected by the information current ${\cal I}_1$.
When we similarly parametrize the changes of the dot level as
\bea
\epsilon_L = \epsilon e^{+\Delta}\,,\qquad
\epsilon_H = \epsilon e^{-\Delta}\,,
\eea
this will for $\Delta \neq 0$ inject energy into the system vie feedback operations.
This secondary type of feedback will not only modify the energy balance (first law), visible in an imbalance between left and right energy currents $I_E^{(L)} \neq - I_E^{(R)}$.
In addition, it also affects the entropic balance via both a modification of the heat flow and the information current ${\cal I}_2$.
These effects are illustrated in Fig.~\ref{FIG:entropic_balance}.

It is clearly visible that neglecting the feedback completely, one may observe an apparent violation of the second law (dashed and solid red curves).
The unconscious injection of energy may lead to a significant increase of the overall produced power (solid red curve) but also implies an apparent violation 
of the second law under Maxwell-demon feedback (solid green curve).
By contrast, the full entropy production rate~(\ref{EQ:entropy_production_specific}) is always positive as expected (black curves).

\subsection{Island charging}\label{SEC:islandcharging}

An additional source for time-dependent rates results from the particles transferred to the metallic island.
For a thermal state~(\ref{EQ:thermalstate}), we can express the total particle number in the metallic island as
$\expval{N_L(t)} = \int {\cal D}_L(\omega) f_L(\omega,t) d\omega$, 
where ${\cal D}_L(\omega)=\sum_k \delta(\omega-\epsilon_{kL})$ denotes its density of states and
$f_L(\omega,t) = [e^{\beta[\omega-\mu_L(t)]}+1]^{-1}$ the Fermi distribution of the metallic island.
When a single electron is added to the metallic island (technically, this can in Eq.~(\ref{EQ:liouvilliantotal}) be tracked with the counting field $\lambda$), 
this will normally displace the reservoir state from thermal equilibrium.
When the coupling to the background reservoir however induces a fast restoration of a thermal equilibrium state immediately thereafter, 
the change of the total particle number is reflected in a change of the chemical potential.
For example, if $\mu_N$ denotes the chemical potential in the island for $N$ particles, we can obtain a relation between two successive
potentials 
\bea\label{EQ:chempotiteration}
1 = \int \left(\frac{{\cal D}_L(\omega)}{e^{\beta(\omega-\mu_{N+1})}+1}-\frac{{\cal D}_L(\omega)}{e^{\beta(\omega-\mu_{N})}+1}\right) d\omega\,.
\eea
For a flat density of states ${\cal D}_L(\omega) = {\cal D}_L^0$ this integral can be solved analytically, 
yielding a simple linear relationship between the chemical potentials $\mu_{N+1} = 1/{\cal D}_L^0+\mu_N$.
This also demonstrates that $e {\cal D}_L^0=e/(\mu_{N+1}-\mu_N)$ is actually the charge capacity of the metallic island.

Throughout this paper we will treat the states in the metallic island as non-interacting, corresponding to a 
flat density of states ${\cal D}_L(\omega)$. 
Inter-island Coulomb interactions would lead to a larger splitting between the higher-excited energy eigenvalues on the island.
Therefore, to model inter-island Coulomb interactions one would have to choose a density of states which 
falls off at large energies, e.g. a Lorentzian one, and we can then solve Eq.~(\ref{EQ:chempotiteration}) numerically for $\mu_{N+1}$. 
One can then check that initially the chemical potential rises linearly with each added charge, 
but as more charges accumulate on the island, its slope increases.
The potential then rises much faster than for a non-interacting island, effectively
suppressing the current even stronger by inhibiting charges from entering the island, compare Eq.~(\ref{EQ:sscur_set}).
Technically, the mean value theorem together with the relation $\int \left[f_1(\omega)-f_2(\omega)\right] d\omega = \mu_1-\mu_2$
ensures that for densities of states with infinite support a solution connecting $\mu_{N+1}$ and $\mu_N$ will always exist.

Finally, we mention that a time-dependent chemical potential will affect the heat currents entering the system as discussed in Sec.~\ref{SEC:entropic_balance}
simply by replacing $\mu\to\mu(t)$, since we have formulated first and second law in a differential way.

\section{Results}\label{SEC:results}

\subsection{Feedback Loop}\label{SEC:feedback}

The counting statistics also yields useful information for the quantum mechanical interpretation of current measurements, and we provide
the formal discussion in App.~\ref{APP:measurements}.
Knowing the statistics of charges traversing the QPC during a measurement time interval $\Delta t$, it becomes straightforward
to simulate a feedback loop on the basis of single trajectories.
Measuring $m$ QPC charges during time interval $\Delta t$ only performs a weak measurement of the SET occupation. 
That means that by measuring e.g. a large time-dependent current $I_m=m/\Delta t$ we cannot be sure that the SET is empty.
From the FCS, we can compute the joint probabilities $P_{nm}(\Delta t)$ of having $-\infty<n<+\infty$ charges 
transferred to the metallic island and $m\ge0$ charges through the QPC (recall that $\gamma_b\to 0$).
Having in mind that the tunneling rates of the SET are much smaller than those of the QPC, the conditional propagator for the relevant processes
can be calculated for $n\in\{-1,0,+1\}$ up to first order in the SET tunneling rates, which enables an efficient numerical simulation of feedback trajectories
in the regime where $\sum_m [P_{-1,m}+P_{0,m}+P_{+1,m}]\approx1$, compare App.~\ref{APP:fcs}.

The simulation of the feedback loop can be summarized as follows:
For each timestep characterized by an SET occupation $n_i$, $N_i$ excess particles on the metallic island, and time-dependent SET tunneling rates
$\Gamma_{L/R}^i$ as well as Fermi functions $f_{L/R}^i$, we compute the 
conditional probabilities $P_{-1,m}(\Delta t)$, $P_{0,m}(\Delta t)$, and $P_{+1,m}(\Delta t)$ of changing the excess particles on the island by $n\in\{-1,0,+1\}$ electrons, 
respectively, and transferring $m\ge 0$ charges through the QPC.
Numerically, one of these possible processes can be selected by using a simple procedure~\cite{press1994}:
Generating a random number $\sigma_1 \in[0,1]$, we select the number of transferred QPC charges $\bar{m}$
that obeys the relation $\sum\limits_{m<\bar{m}} (P_{-1,m}+P_{0,m}+P_{+1,m}) < \sigma_1\allowbreak < \sum\limits_{m\le\bar{m}} (P_{-1,m}+P_{0,m}+P_{+1,m})$.
In a similar fashion, one can then determine the particle number $\bar{n}$ exchanged with the metallic island from comparing a random number $\sigma_2\in[0,1]$
with the conditional probabilities 
$\sum\limits_{n<\bar{n}} \frac{P_{n,\bar{m}}}{P_{\bar{m}}} < \sigma_2 \allowbreak < \sum\limits_{n\le\bar{n}} \frac{P_{n,\bar{m}}}{P_{\bar{m}}}$.
With the specific trajectory $(\bar{n},\bar{m})$ chosen, we can update the state of the SET (compare App.~\ref{APP:measurements}), update the island excess charge, and also determine the 
time-dependent QPC current, respectively, 
\bea
\rho_{i+1} &\propto& \frac{1}{(2\pi)^2} \int_{-\pi}^{+\pi} d\lambda d\xi e^{{\cal L}(0,\lambda,\xi)\Delta t} e^{-\ii \bar{n} \lambda} e^{-\ii \bar{m} \xi} \rho_i\,,\nn
N_{i+1} &=& N_i + \bar{n}\,,\qquad
I_i = \bar{m}/\Delta t\,.
\eea
To close the feedback loop, we condition the tunneling rates in the next interval on the measurement outcome.
Defining a discrimination threshold $m_{\rm th} = \gamma_f \Delta t(1+\kappa)/2$ as before, we choose 
\bea
\Gamma_{\alpha}^{i+1} &=& \left\{\begin{array}{ccc}
\Gamma_{\alpha}^H & : & m \ge m_{\rm th}\\
\Gamma_{\alpha}^L & : & m < m_{\rm th}
\end{array}\right.\,,\nn
f_\alpha^{i+1} &=& \left\{\begin{array}{ccc}
f_\alpha^H & : & m \ge m_{\rm th}\\
f_\alpha^L & : & m < m_{\rm th}
\end{array}\right.\,.
\eea
This automatically includes that the feedback is sometimes erroneous.

\begin{figure}[t]
\includegraphics[width=0.48\textwidth,clip=true]{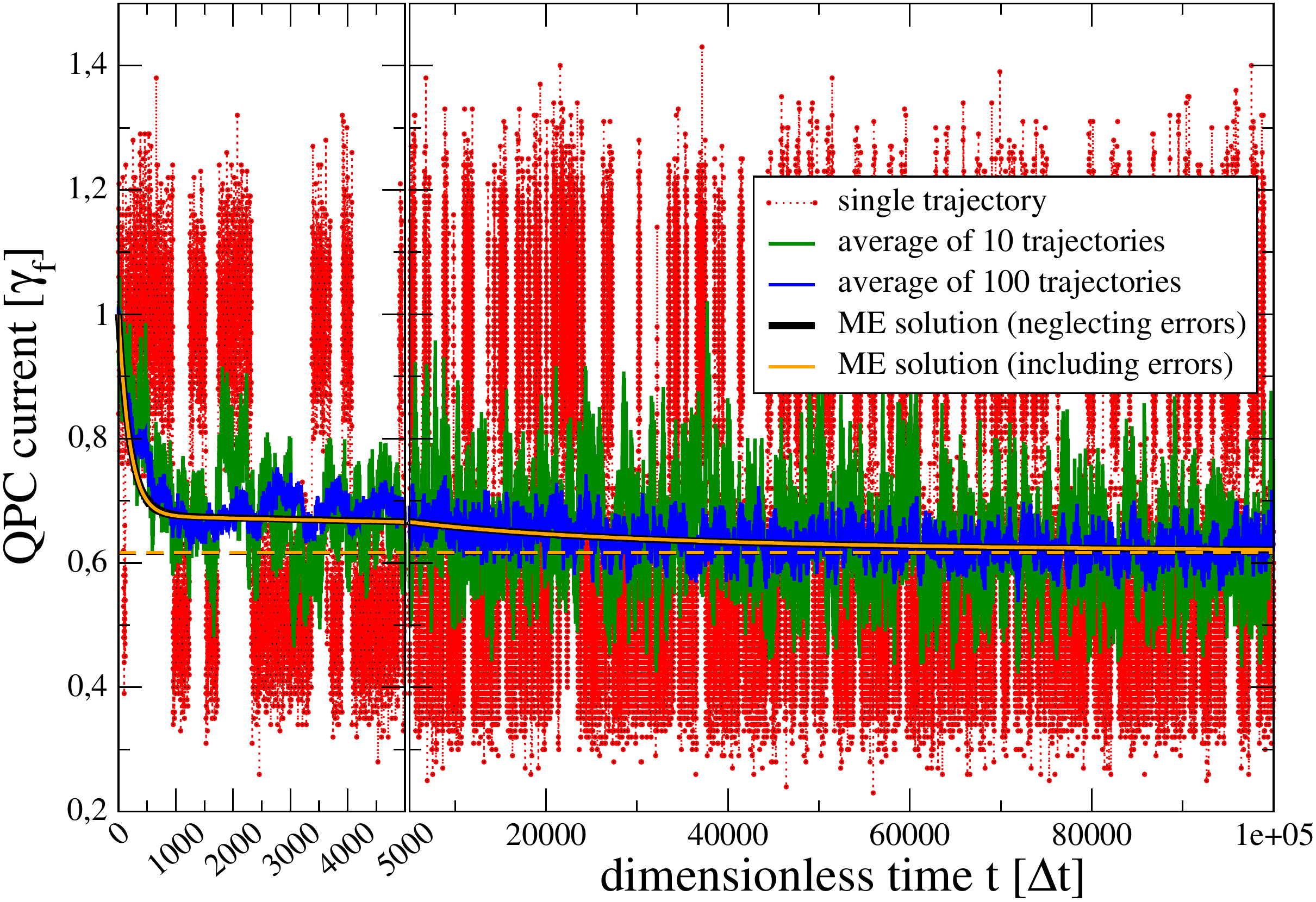}
\caption{\label{FIG:current_trajectories}
Plot of the time-dependent QPC current for a single trajectory (symbols and dotted red) and averages of 10 (green) and 100 (blue) trajectories, all
starting with an initially empty SET.
The discrimination threshold is chosen as the average of the two mean currents $I_{\rm discr} = 0.75 \gamma_f$, and feedback is for each trajectory conditioned on the
QPC current.
The average evolutions (solid black and orange, compare Sec.~\ref{SEC:average_system} and Sec.~\ref{SEC:average_potential}) first show a rapid decay (left) and then slowly relax (right) toward the 
true steady state value (dashed) on a much longer timescale.
Parameters: $\Gamma_L^H = \Gamma_R^L = 0.1 \Gamma$, $\Gamma_L^L = \Gamma_R^H = 0.9 \Gamma$, $\beta \Gamma = 0.01$, $\epsilon^H=\epsilon^L = \epsilon$, $\beta\epsilon=1$, 
$\beta \mu_L^0 = 3$, $\beta \mu_R = 0$, ${\cal D}_L^0/\beta = 10$, $\kappa=0.5$, $\gamma_f \Delta t=100$.
The average detection error~(\ref{EQ:detection_error}) is slightly time-dependent but small $P_{\rm err} \le .217\%$.
}
\end{figure}

\begin{figure}[t]
\includegraphics[width=0.48\textwidth,clip=true]{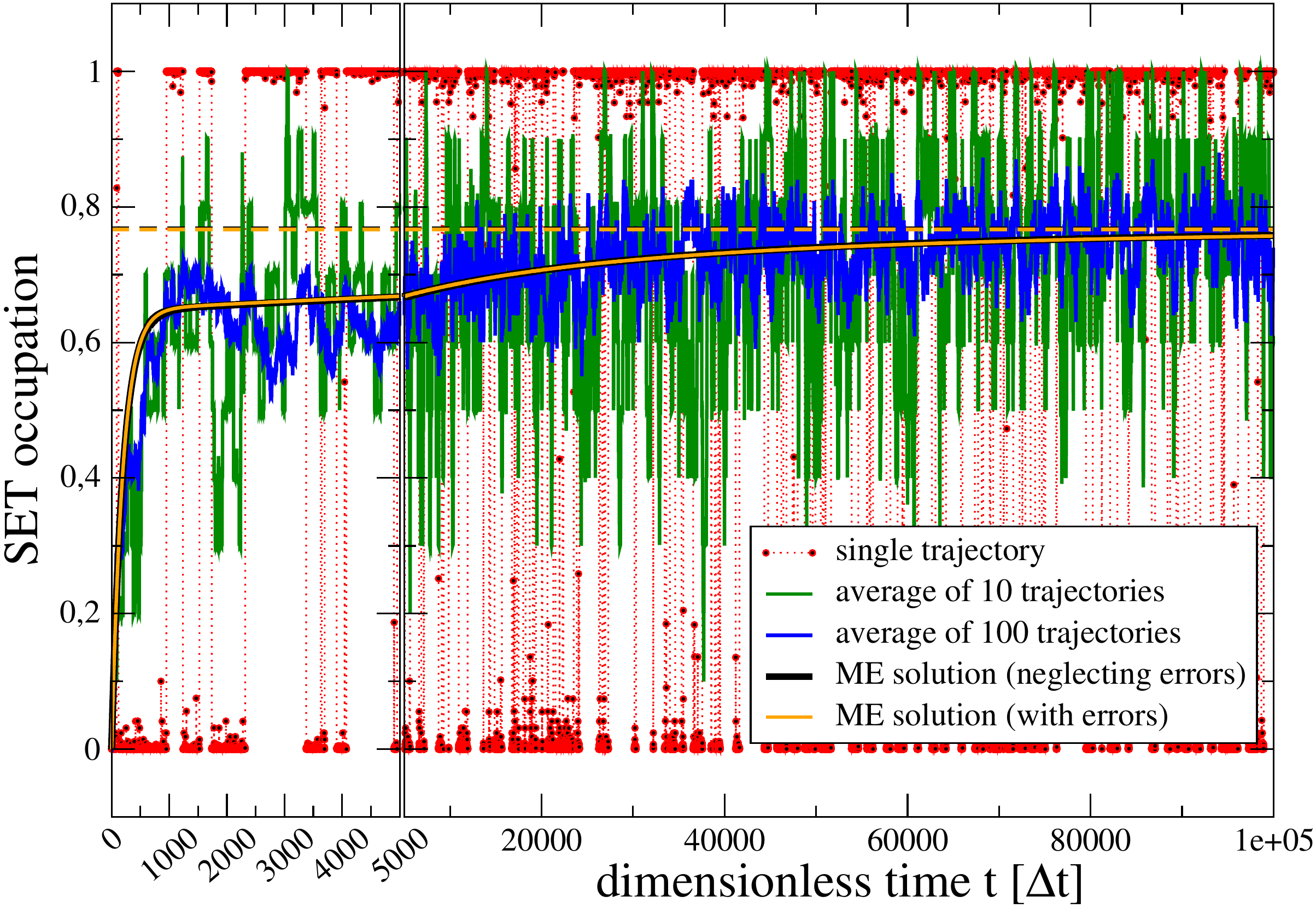}
\caption{\label{FIG:occupation_trajectories}
Plot of the time-dependent SET occupation for a single trajectory and averages of multiple trajectories.
Color coding and parameters have been chosen as in Fig.~\ref{FIG:current_trajectories}.
}
\end{figure}

\begin{figure}[t]
\includegraphics[width=0.48\textwidth,clip=true]{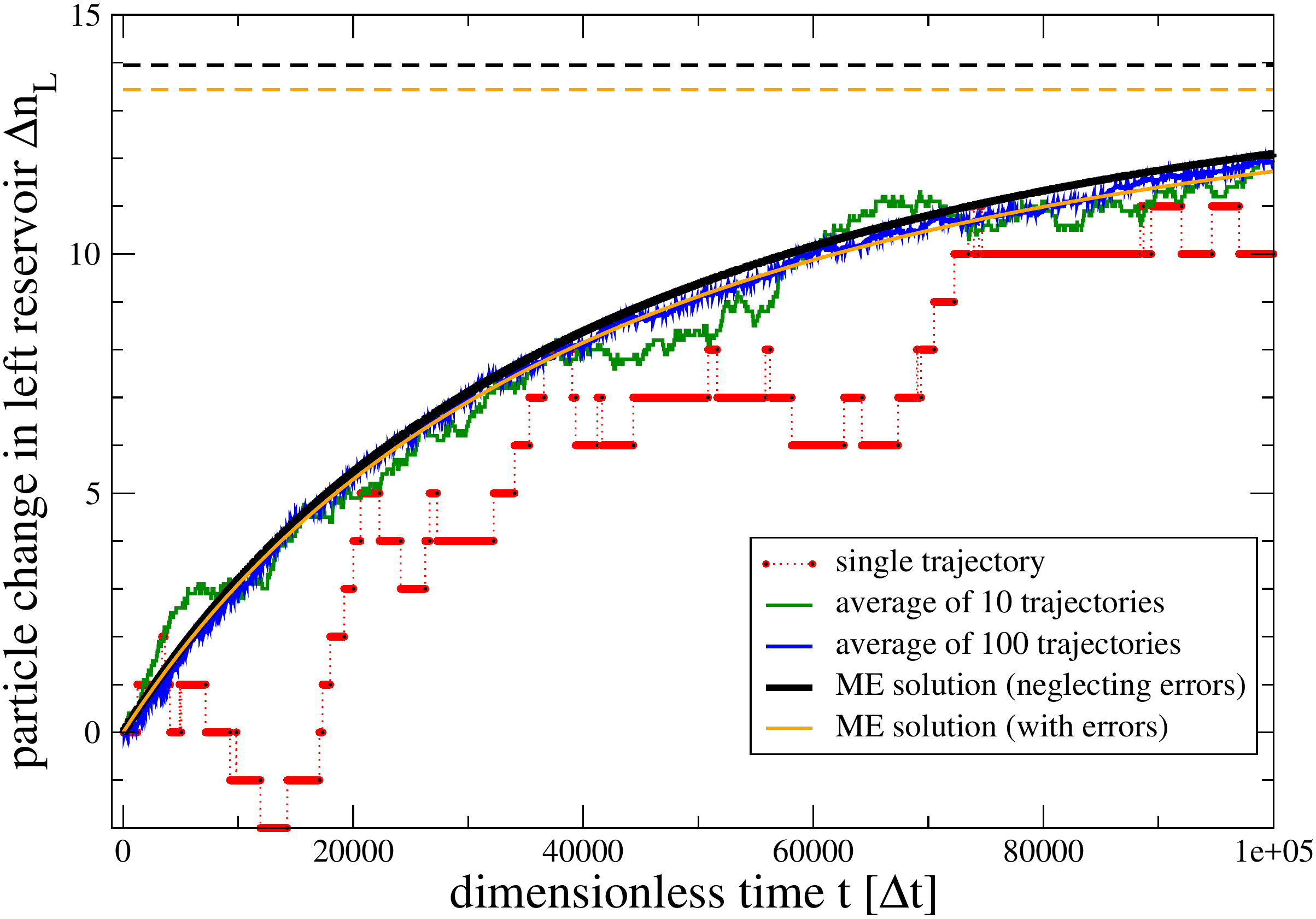}
\caption{\label{FIG:pnumber_trajectories}
Plot of the time-dependent occupation of the metallic island for a single trajectory and averages of multiple trajectories.
The neglect of measurement error has a non-negligible impact on the average evolution (solid orange and black), and the average of trajectories 
fits much better when the measurement error is included.
Color coding and parameters have been chosen as in Fig.~\ref{FIG:current_trajectories}.
}
\end{figure}

With this feedback loop, we can model the time-dependent evolution of the QPC current, the
SET occupation, and the excess population in the metallic island, see Fig.~\ref{FIG:current_trajectories}, 
Fig.~\ref{FIG:occupation_trajectories}, and Fig.~\ref{FIG:pnumber_trajectories}.

A low QPC current can be associated with an empty SET and would erroneously lead to the wrong feedback operation.
However, for the chosen parameters we see that these events are rather unlikely.
This is illustrated in Fig.~\ref{FIG:occupation_trajectories}.
In particular when the QPC current observed during $\Delta t$ is inconclusive, the SET occupation is not fully projected
to empty or filled but assumes some intermediate value.
However, also here we see after a very fast initial relaxation in the first few hundred iterations a much slower increase
of the average SET occupation. 
This goes along with a slow increase of the chemical potential in the metallic island.

This change of the potential in the metallic island is directly linked to its excess particle number, see Fig.~\ref{FIG:pnumber_trajectories}.
As the feedback is constructed to pump electrons into the metallic island, its particle content will -- on average -- increase.
Individual trajectories may however also show a temporary decrease of the metallic island charge occupation.

Here, we see that the stationary configuration in the chemical potential is not reached in the observed timeframe, but the
QPC current and the SET occupation are not so sensitive in this regime.
For interactions present within the metallic island, one would expect the potential to reach its maximally charged value on even larger timescales, 
as the current is even stronger suppressed.

\subsection{Average evolution under feedback}\label{SEC:average_system}

Many idealized descriptions of feedback-control circuits separate
the measurement and control operations in time~\cite{sagawa2008a,jacobs2009a,schaller2012a}.
Here, the situation is slightly different as the information determining the future evolution is gathered over the
complete timestep $\Delta t$.
When both the measurement of the QPC particle number and the feedback control operations occur instantaneously,
the density matrix at time $t+\Delta t$ is related to the density matrix at time $t$ via
\bea
\rho^{(m)}(t+\Delta t) = e^{{\cal L}_m \Delta t} {\cal M}_m(\Delta t) \rho(t)/P_m\,.
\eea
Here, ${\cal M}_m(\Delta t)$ describes the measurement superoperator associated to measurement outcome $m$, 
$P_m=\trace{{\cal M}_m (\Delta t) \rho(t)}$ the corresponding probability, and ${\cal L}_m$ the conditional evolution.
The measurement superoperators ${\cal M}_m (\Delta t)$ are assumed to imply an instantaneous collapse but depend 
on the sampling time $\Delta t$.
Without feedback, we discuss how they can be obtained from the FCS of the QPC in App.~\ref{APP:measurements}.
Performing a weighted average over all outcomes, we obtain
\bea\label{EQ:fixed_point}
\rho(t+\Delta t) = \left[\sum_m e^{{\cal L}_m \Delta t} {\cal M}_m(\Delta t)\right] \rho(t)\,,
\eea
which can be expanded for small $\Delta t$ in the continuum measurement limit.
We aim to obtain an effective feedback Liouvillian ${\cal L}_{\rm fb}$ acting on the SET only.

To obtain the measurement superoperator ${\cal M}_m \rho(t) \equiv \rho^{(m)}(t)$, we recall that the conditional
density matrix can be obtained via an inverse Fourier transform
\bea
\rho^{(m)}(t) &=& \frac{1}{2\pi} \int_{-\pi}^{+\pi} \rho(\xi,t) e^{-\ii m \xi} d\xi\\
&=& \frac{1}{2\pi} \int_{-\pi}^{+\pi} e^{{\cal L}_{\rm fb}\Delta t + {\cal L}_{\rm dt}(\xi)\Delta t-\ii m \xi} d\xi \rho(t-\Delta t)\,,\nonumber
\eea
where we have postulated the existence of an a priori unknown effective SET feedback Liouvillian.
We also see that the information about the measurement has been collected in the 
previous timestep during $[t-\Delta t,t]$.
Inserting the backward evolution ignorant of the QPC charges $\rho(t-\Delta t)=e^{-{\cal L}_{\rm fb} \Delta t} \rho(t)$, we
can infer for the measurement superoperator
\bea
{\cal M}_m(\Delta t) = \frac{1}{2\pi} \int\limits_{-\pi}^{+\pi} e^{{\cal L}_{\rm fb}\Delta t + {\cal L}_{\rm dt}(\xi)\Delta t-\ii m \xi} d\xi e^{-{\cal L}_{\rm fb}\Delta t}\,.
\eea
Although corresponding to a projective measurement in the enlarged SET-detector Hilbert space, in the reduced
SET Hilbert space these measurements appear only as a weak measurement, exemplifying Neumarks theorem~\cite{peres1990a}.
In particular, we find that $\sum_m {\cal M}_m(\Delta t) = \f{1}$, which enables to derive from Eq.~(\ref{EQ:fixed_point}) in the
continuum measurement limit ($\Delta t \to 0$)
a master equation $\dot{\rho} = {\cal L}_{\rm fb} \rho$ with an effective feedback Liouvillian determined self-consistently by the equation
\bea\label{EQ:lmatfb0}
{\cal L}_{\rm fb} = \sum_m {\cal L}_m {\cal M}_m(\Delta t)\,.
\eea
As there are just two conditional SET dissipators  
${\cal L}^H$ and ${\cal L}^L$, one for a high and one for a low QPC current 
(ideally corresponding to an empty and filled dot, respectively), the effective Liouvillian becomes
${\cal L}_{\rm fb} = {\cal L}^H {\cal P}^H + {\cal L}^L {\cal P}^L$, 
where the measurement superoperators for a high/low current, respectively, become
\bea
{\cal P}^H &=& \sum_{m\ge m_{\rm th}} {\cal M}_m\,,\qquad {\cal P}^L = \sum_{m < m_{\rm th}} {\cal M}_m\,.
\eea
The calculation of these measurement superoperators will be generally difficult and needs to be performed numerically.

However, considering the limit of interest where $\gamma_f \Delta t \gg 1$ while ${\cal L}_{\rm fb}^{01/10} \Delta t \ll 1$, we see that 
we can obtain an approximate expression for the measurement superoperators
\bea
{\cal M}_m &\approx& \frac{1}{2\pi} \int\limits_{-\pi}^{+\pi} e^{{\cal L}_{\rm dt}(\xi)-\ii m \xi} d\xi\\
&=& \frac{1}{m!}\left(\begin{array}{cc}
e^{-\gamma_f \Delta t} (\gamma_f \Delta t)^m & 0\\
0 & e^{-\kappa \gamma_f \Delta t} (\kappa \gamma_f \Delta t)^m
\end{array}\right)\,,\nonumber
\eea
such that we can infer -- compare Eqns.~(\ref{EQ:det_errors}) -- for the measurement superoperators corresponding to high and low QPC currents the relation
\bea\label{EQ:measurement1}
{\cal P}^L &=& \frac{1}{\Gamma(1+m_{\rm th})}\times\nn
&&\times \left(\begin{array}{cc}
\Gamma(1+m_{\rm th}, \gamma_f \Delta t) & 0\\
0 & \Gamma(1+m_{\rm th}, \kappa \gamma_f \Delta t)
\end{array}\right)\,,\nn
{\cal P}^H &=& \f{1}-{\cal P}^L\,.
\eea
Indeed one can show that for $\kappa<1$ and properly chosen discrimination threshold 
$\kappa \gamma_f \Delta t < m_{\rm th} < \gamma_f \Delta t$
these superoperators approach the idealized projectors when $\gamma_f \Delta t \to \infty$
\bea\label{EQ:measurement2}
{\cal P}^H \to {\cal P}^E = \left(\begin{array}{cc}
1 & 0\\
0 & 0
\end{array}\right)\,,\qquad
{\cal P}^L \to {\cal P}^F = \left(\begin{array}{cc}
0 & 0\\
0 & 1
\end{array}\right)\,.
\eea

We can now compute the effective feedback Liouvillian and derive from it 
the current $I_M^{(L)}$ entering the system from the metallic island.

\subsection{Average potential evolution under feedback}\label{SEC:average_potential}

The current in turn determines the evolution of the chemical potential on the island 
via (similar to Sec.~\ref{SEC:islandcharging})
\bea\label{EQ:diffeq_chempot}
\expval{\dot{N}_L(t)} &=& \int {\cal D}_L(\omega) f_L(\omega,t) [1-f_L(\omega, t)] d\omega \beta \dot{\mu}_L\nn
&=& D_L^0 \dot{\mu}_L = -I_M^{(L)}(t)\,,
\eea
where we have again assumed a constant density of states ${\cal D}_L(\omega) = {\cal D}_L^0$.
Since the current depends on the value of the chemical potential of the island -- compare e.g. the steady-state solution
in Eq.~(\ref{EQ:sscur_set}) -- this defines a nonlinear differential equation for $\mu_L(t)$, which can be solved numerically.

\subsubsection{Neglecting measurement errors}

The neglect of measurement errors corresponds to the effective feedback Liouvillian
${\cal L}_{\rm fb} = {\cal L}^L {\cal P}^F + {\cal L}^H {\cal P}^E$, which becomes explicitly
Eq.~(\ref{EQ:lmatfb}).
This only extends on previous results~\cite{schaller2011b} by allowing for the possibility of
changing the dot level -- marked by an index of the Fermi functions.
The associated current entering the system from the metallic island reads
\bea\label{EQ:fbcurrent1}
I_M^{(L)} = \frac{\Gamma_L^H \Gamma_R^L f_L^H (1-f_R^L)-\Gamma_L^L \Gamma_R^H (1-f_L^L) f_R^H}
{\Gamma_L^H f_L^H + \Gamma_L^L (1-f_L^L)+\Gamma_R^H f_R^H + \Gamma_R^L (1-f_R^L)}\,.\nn
\eea
It depends implicitly on $\mu_L(t)$ via the left-associated Fermi functions.
This yields a steady-state relation for $\mu_L(t)$, and in the simplified case where the
feedback is purely informational~\cite{esposito2012a} (Maxwell-demon limit, $f_\alpha^H=f_\alpha^L=f_\alpha$), we obtain 
for the island potential now the steady state value
\bea\label{EQ:sschempot1}
\bar\mu_L = \mu_R + \frac{1}{\beta} \ln \frac{\Gamma_L^L \Gamma_R^H}{\Gamma_L^H \Gamma_R^L}\,.
\eea
Here, one can directly see that choosing $\Gamma_L^L$ and $\Gamma_R^H$ large and $\Gamma_L^H$ and $\Gamma_R^L$ small
will raise the chemical potential of the island above the right lead potential, which is also the protocol chosen
in Figs.~\ref{FIG:current_trajectories},~\ref{FIG:occupation_trajectories}, and~\ref{FIG:pnumber_trajectories}.
Furthermore, we also see that finite temperatures a necessary prerequisite to raise the island potential
as expected for a Maxwell demon protocol.
The solid black and dashed black curves in these figures correspond to the error-free feedback limit, 
and resemble the average current and SET occupations well.

\subsubsection{Including measurement errors}

With choosing ${\cal L}_{\rm fb} = {\cal L}^L {\cal P}^L + {\cal L}^H {\cal P}^H$ we also consider feedback errors.
The measurement superoperators are no longer projectors, and the explicit form of the feedback
Liouvillian becomes more complicated (not shown).
In the entropic balance, it is much more complicated to differentiate between informational feedbacks and energy-injecting feedbacks.
The general procedure for the current however is identical to the previous subsection, and the results for time-dependent
SET occupation, QPC current, and island population will be slightly changed.
For example, in the limit of purely informational feedback ($f_\alpha^L=f_\alpha^H=f_\alpha$), the steady-state potential of the island becomes
\bea\label{EQ:sschempot2}
\bar\mu_L &=& \mu_R + \frac{1}{\beta} \times\\
&&\ln 
\frac{\left[\left(\Gamma_L^L - \Gamma_L^H\right) g_2 + \Gamma_L^H\right]\left[\Gamma_R^H + \left(\Gamma_R^L-\Gamma_R^H\right) g_1\right]}
{\left[\Gamma_L^H + \left(\Gamma_L^L-\Gamma_L^H\right) g_1\right]\left[\left(\Gamma_R^L-\Gamma_R^H\right)g_2 + \Gamma_R^H\right]}\nn
g_1 &=& \frac{\Gamma(1+m_{\rm th}, \gamma_f \Delta t)}{\Gamma(1+m_{\rm th})}\,,\;\;
g_2 = \frac{\Gamma(1+m_{\rm th}, \kappa \gamma_f \Delta t)}{\Gamma(1+m_{\rm th})}\,.\nonumber
\eea
In the appropriate limit of an error-free measurement, we have $g_1\to 0$ and $g_2\to 1$, such that Eq.~(\ref{EQ:sschempot1}) 
is recovered.
When the measurement is insensitive ($g_1=g_2$), it is not possible to charge the island at all and we have $\bar\mu_L=\mu_R$.
When we choose $\kappa > 1$, the measurement will always yield the wrong result, we have $g_1\to 1$ and $g_2\to 0$, and the 
potential is dragged into the opposite direction, just as if the opposite feedback protocol was chosen.
For specific finite $0\le g_i \le 1$ consistent with the chosen parameters, the results are depicted with the solid and dashed orange curves in
Figs.~\ref{FIG:current_trajectories},~\ref{FIG:occupation_trajectories}, and~\ref{FIG:pnumber_trajectories}.
Whereas the measurement errors hardly have an effect on the SET occupation and QPC current, their effect on the
evolution of the island potential or, equivalently, the island population in Fig.~\ref{FIG:pnumber_trajectories} 
is more pronounced.
Here, also the steady-state population is significantly reduced by erroneous measurements, and as expected, measurement
errors tend to reduce the effect of the feedback.

\section{Discussion}

\subsection{Remarks on experiments}\label{SEC:experiments}

To reconstruct the original Maxwell-demon scenario, it is not desirable to drive the dot levels. 
However, by changing the potentials of the gate controls this will in reality usually happen.
Therefore, it is advisable to apply the two Liouvillians ${\cal L}^L$ and ${\cal L}^H$ alternatingly, i.e., 
in an unconditioned turnstyle with equal duration.
One can show for the particular model that in equilibrium ($\mu_L=\mu_R$ such that $f_\alpha^H \to f^H$ and $f_\alpha^L \to f^L$) 
the average electronic current only vanishes for vanishing driving (when $f^H=f^L$).
This is the reverse of the no-pumping theorem~\cite{rahav2008a} and gives an experimental recipe for fine-tuning 
the control protocol to minimize the energy injection due to the feedback.

Before experimental data can be interpreted, it is necessary to determine the microscopic parameters in the master equation.
Here, we just sketch how some of these can be obtained by statistical analysis of the QPC current data.

Obviously, the two mean values of the current may serve to determine $\gamma_f$ and $\kappa$.
Furthermore, one can use the width of these time-dependent currents (compare shaded regions in Fig.~\ref{FIG:trajectory_example})
to see whether the statistics is really unimodal Poissonian or whether there are additional contributions resulting from a finite backward transition 
rate $\gamma_b$.

The mean, variance, and higher cumulants of the statistics of total SET jumps can be related to the SET parameters, 
and it is an experimentally well-established procedure to extract even very large cumulants~\cite{flindt2009a} from QPC data.
Theoretically, they can be extracted using the Full Counting Statistics exposed in App.~\ref{APP:fcs}.
For example, the mean rate of jumps is given by 
\bea
\expval{\dot{n}_{\rm jp}} = 2 \frac{{\cal L}_{\rm fb}^{01} {\cal L}_{\rm fb}^{10}}{{\cal L}_{\rm fb}^{01}+{\cal L}_{\rm fb}^{10}}\,,
\eea
and it is straightforward to compute higher moments or cumulants.

Similarly, one can obtain the waiting time distributions for an empty or filled SET, respectively
\bea
{\cal P}_E(\tau) = {\cal L}_{\rm fb}^{10} e^{-{\cal L}_{\rm fb}^{10} t}\,,\qquad
{\cal P}_F(\tau) = {\cal L}_{\rm fb}^{01} e^{-{\cal L}_{\rm fb}^{01} t}\,,
\eea
from which one can separately extract the transition rates under feedback by computing the corresponding mean and variances.

Unfortunately, these quantities are only sensitive to the total rates and cannot resolve between right- and left-associated processes.
As the transition rates are related microscopically to the feedback rates via 
${\cal L}_{\rm fb}^{10} = \Gamma_L^H f_L^H + \Gamma_R^H f_R^H$ and ${\cal L}_{\rm fb}^{01} = \Gamma_L^L (1-f_L^L) + \Gamma_R^L (1-f_R^L)$,
their evolution could be compared in time to judge whether the island has indeed been charged.
It will be necessary to compare the statistics with different feedback protocols and also without feedback to obtain the parameters in the Liouvillian. 

Most optimal however would be an additional witness of the island charging predicted in Eqns.~(\ref{EQ:sschempot1}) and~(\ref{EQ:sschempot2}).
This could be implemented by using an additional QPC that only monitors the island.
In reality however, the QPC present in the current setup could also be sensitive to the charge accumulated on the island.
Technically, this could be modeled by an additional interaction Hamiltonian $H_{L, \rm dt}$ in Eq.~(\ref{EQ:hams}), 
which would lead to a slow time-dependent drift in both mean values of the QPC currents as charge accumulates on the island.

\subsection{Other applications}\label{SEC:applications}

We have mainly discussed the device with a focus on its operation as a Maxwell demon that is capable of working against 
an electric potential by assuming the leads at the same temperature.
However, even in absence of feedback, it is conceivable to use other operational modes, see also Ref.~\cite{esposito2012a}.

For example, when applying a temperature gradient between the right lead (with a fixed chemical potential) 
and the background reservoir of the metallic island, the island potential will evolve until the currents vanish, 
compare Eq.~(\ref{EQ:fbcurrent1}) in the limit $\Gamma_\alpha^{H/L} \to \Gamma_\alpha$ and $f_\alpha^{H/L} \to f_\alpha$.
Thus, a temperature gradient induces a voltage difference between island and right lead (Seebeck effect).
Feedback can be used to enhance or reduce this.

On the other hand, one may also exploit the Peltier effect to cool the island (and its background reservoir).
As a transient effect, this would require dot level and initial island potential to be tuned such that $\dot{Q}^{(L)}=(\epsilon-\mu_L^0) I_M>0$ 
(neglecting level shifts due to feedback).
For a finite-sized island, this will be transient as the potential will evolve until the current vanishes.
However, with feedback active, one may significantly enhance the heat current by pumping electrons out of the metallic island.
In effect, this will lead to a cooling of the island (and its background reservoir).

\subsection{Summary}\label{SEC:summary}

We have considered an external feedback loop on a single electron transistor connecting a metallic
island with a time-dependent potential and charge and a lead with a fixed potential.
The feedback was conditioned on the current measured by a nearby QPC and consisted of piecewise-constant
switching of the bare tunneling rates.
As the current does not always faithfully reflect the SET occupation, the inclusion of a detector device into the
description automatically yields a microscopic error model and could thus aid in the design of effective error models~\cite{wagner2016a}.
With an appropriate feedback protocol chosen, we can charge the metallic island -- corresponding to a raising its potential
above the lead level.
Ideally, this is done using only the information from the QPC and would thus correspond to an electronic implementation
of a Maxwell demon.
However, the control operations may also lead to a turnstyle switching of the dot level, 
thereby injecting energy into the system and departing from the ideal Maxwell demon limit.
We have discussed the entropic balances for ideal Maxwell-demon feedbacks~\cite{esposito2012a} 
and energy-injecting feedbacks, and the modification due to these will have to be taken into account in
an experiment.

Most important, we have seen that although small, measurement errors may have some impact on the achieved charging of the island
and should be considered in an actual experiment.

Studying the additional influence of delay between measurement and control (compare e.g.~\cite{emary2013a}),
comparing the demon performance with electronic pumps (compare e.g.~\cite{juergens2013a}),
and investigating how feedback affects other operational modes (e.g. the efficiency of thermoelectric generators or the coefficient of
performance of heat pumps) are interesting future research perspectives.

\begin{acknowledgement}
The author gratefully acknowledges financial support by the DFG (SCHA 1646/3-1, SFB 910, GRK 1558) and very helpful discussions with
T. Brandes, M. Esposito, R. Haug, and T. Wagner.
\end{acknowledgement}

\bibliographystyle{pss}
\providecommand{\WileyBibTextsc}{}
\let\textsc\WileyBibTextsc
\providecommand{\othercit}{}
\providecommand{\jr}[1]{#1}
\providecommand{\etal}{~et~al.}


\begin{appendix}

\section{Full Counting Statistics}\label{APP:fcs}

For the time-dependent counting statistics the full propagator
${\cal P}(\f{\chi}, \Delta t)\allowbreak=e^{{\cal L}(\f{\chi}) \Delta t}$ is a central object, which can be
understood with a few examples:

First of all, in absence of feedback and for an initial state $\rho_0$, the moment-generating function for the 
probability distributions of total jumps observed, net charges transferred to the left junction, 
and net charges transferred through the QPC detector is given by ${\cal M}(\f{\chi},t) = \trace{{\cal P}(\f{\chi}, t) \rho_0}$.
Computing derivatives with respect to the counting fields yields the moments of the distribution, e.g. for the mean number
of total jumps one has to evaluate $\expval{n} = \left.(-\ii \partial_\chi) {\cal M}(\f{\chi},t)\right|_{\f{\chi}\to\f{0}}$.

The actual probabilities have to be obtained via inverse Fourier transform.
For example, to obtain the probability $P_n^{\rm QPC}(\Delta t)$ of transferring $n$ charges through the QPC during the
time interval $\Delta t$, one has to evaluate the integral
\bea
P_n^{\rm QPC}(\Delta t) = \frac{1}{2\pi} \int_{-\pi}^{+\pi} {\cal M}(0,0,\xi,\Delta t) e^{-\ii n \xi} d\xi\,.
\eea

Furthermore, upon actually measuring $n$ QPC particles during $\Delta t$, the system density matrix changes according to (see also App.~\ref{APP:measurements})
\bea\label{EQ:qpcevolution}
\rho(t+\Delta t) &\stackrel{n}{\to}& \frac{{\cal J}^{(n)}(\Delta t)\rho(t)}{P_n^{\rm QPC}(\Delta t)}\,,\nn
{\cal J}^{(n)}(\Delta t) &=& \frac{1}{2\pi} \int_{-\pi}^{+\pi}{\cal P}(0,0,\xi,\Delta t) e^{-\ii n \xi} d\xi\,.
\eea
A trivial limit arises when the QPC is not sensitive to the SET state ($\kappa=1$).
Then, the Liouvillians ${\cal L}_{\rm dt}$ and ${\cal L}_L+{\cal L}_R$ commute, and QPC and SET evolve independently from each other. 
In the unidirectional QPC limit ($\gamma_b \to 0$), the conditional propagator then becomes a product of unchanged SET evolution multiplied by a Poissonian probability distribution
\bea
{\cal J}^{(n)}(\Delta t) = e^{({\cal L}_L+{\cal L}_R) \Delta t} \frac{(\gamma_f \Delta t)^n}{n!} e^{-\gamma_f \Delta t}\,.
\eea

\begin{figure}[ht]
\includegraphics[width=0.5\textwidth,clip=true]{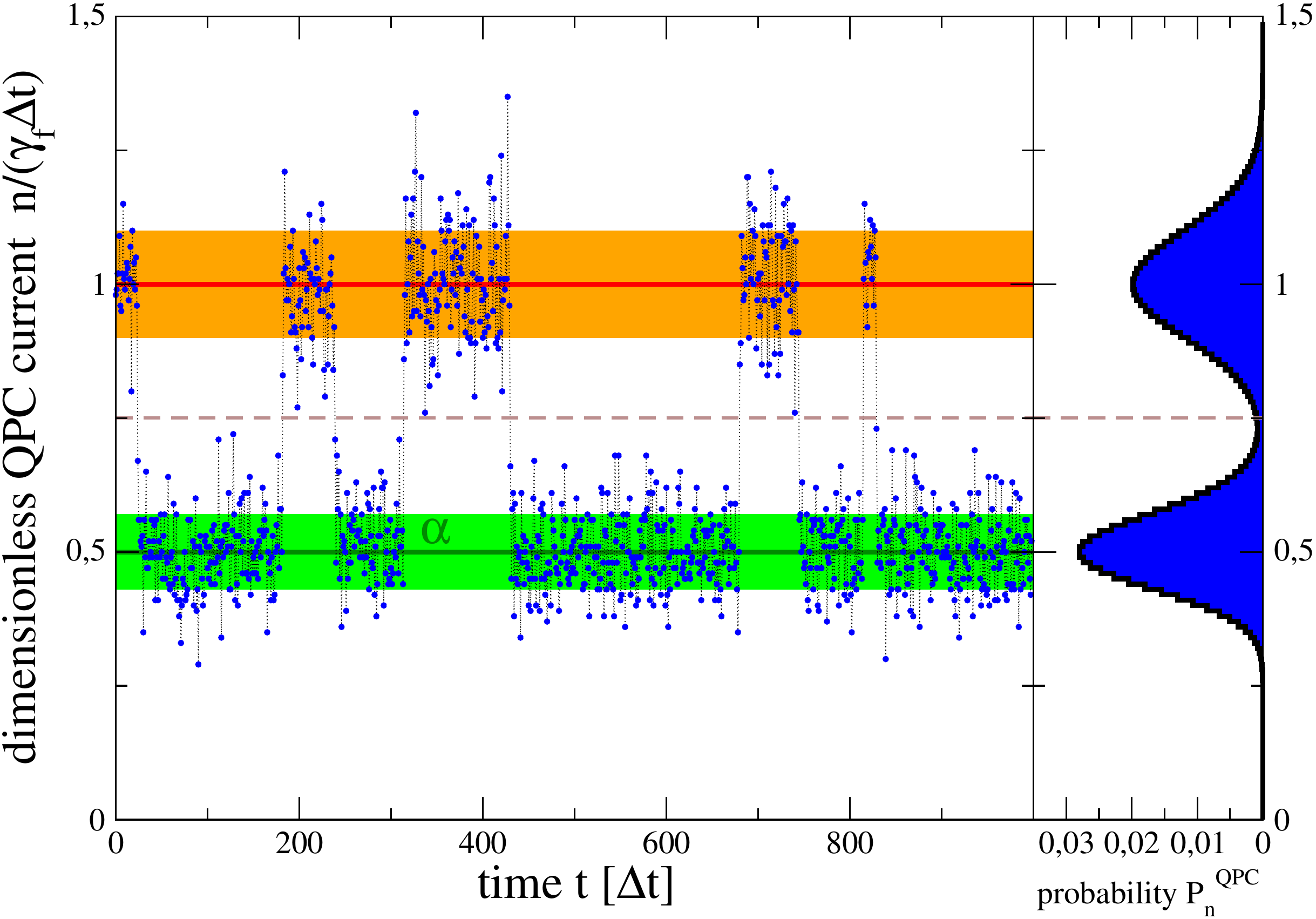}
\caption{\label{FIG:trajectory_example}
Plot of a randomly generated QPC current trajectory based on Eq.~(\ref{EQ:qpcevolution}).
Points represent measured QPC currents, which is expected at $\gamma_f$ when the SET is empty and 
at $\kappa \gamma_f$ when the SET is filled.
Thick lines and shaded regions denote mean and standard deviation of the Poissonian process described by 
$\lambda_E = \gamma_f \Delta t$
and $\lambda_F = \kappa \gamma_f \Delta t$ for an empty (red and orange) or filled (dark and light green) SET, 
respectively, whereas the distribution on the right displays
the full bimodal distribution obtained for an infinitely long trajectory (but at finite $\Delta t$).
The deliberately chosen discrimination threshold at $(1+\kappa)/2$ (dashed) clearly separates the two peaks on the right 
(a slightly smaller threshold would be more optimal).
Parameters ${\cal L}^{01}\Delta t = {\cal L}^{10}\Delta t = 0.01$, $\gamma_f \Delta t = 100$, $\kappa=1/2$, such that the
error per measurement becomes $P_{\rm err} \approx 0.294\%$, compare Eq.~(\ref{EQ:detection_error}).
}
\end{figure}

In the general case ($\kappa \neq 1$), the propagator becomes significantly more sophisticated.
Eq.~(\ref{EQ:qpcevolution}) enables us to implement an iteration scheme for the density matrix, 
which also yields the time-dependent QPC currents $I_n = n/\Delta t$,
see Fig.~\ref{FIG:trajectory_example}.

To enable analytic approximations, we note that the Laplace transform of the propagator 
$\int_0^\infty {\cal P}(\f{\chi}, t) e^{-z t} dt$ can be analytically determined 
and enables the calculation of the first few moments of all desired probabilities.
It may also aid the direct calculation of the probabilities by computing a series expansion:
In the limit where the QPC can be used as a charge detector, we require that its current is macroscopic
in comparison to the charge throughput of the SET, i.e., formally that $\gamma_{f} \gg \Gamma_{\alpha}$.
Generally, we can perform an expansion of the propagator when we split the Liouvillian as ${\cal L} = {\cal L}_0 + {\cal L}_1$ for small ${\cal L}_1$, 
where we obtain for the Laplace transform of the propagator
\bea
{\cal P}(z) &=& \frac{1}{z\f{1}-{\cal L}_0-{\cal L}_1} = \sum_{n=0}^\infty \left[{\cal P}_0(z) {\cal L}_1\right]^n {\cal P}_0(z)\,,\nn
{\cal P}_0(z) &=& \frac{1}{z\f{1}-{\cal L}_0}\,.
\eea
Since we know that the SET tunneling rates are much smaller than the QPC tunneling rates, we choose 
${\cal L}_0 = {\cal L}_{\rm dt}$ and the perturbation ${\cal L}_1 = {\cal L}_L+{\cal L}_R$.
The propagator for the unperturbed evolution ${\cal P}_0(z)$ can be calculated explicitly, 
which enables to calculate the SET-induced first order correction (and also higher orders if desired).
Most importantly, we can now calculate the first correction by evaluating the inverse Laplace transform of
${\cal P}_0(\xi,z) {\cal L}_1 {\cal P}_0(\xi,z)$.
Alternative splittings could arise from an expansion around the interaction-free limit ($\kappa=1$).

Finally, to obtain the joint probability of e.g. a particle transfer of $n$ charges to the left metallic 
island and $m$ charges through the QPC, we have to evaluate the propagator ${\cal P}(0, \chi, \xi, \Delta t)$.
Below, we demonstrate the usefulness of the joint Full Counting Statistics by calculating an error estimate for the
QPC detector.

\section{Detection Errors}\label{APP:errors}

The usefulness of the QPC as a charge detector crucially depends on its sampling frequency, characterized by
the time interval $\Delta t$ between two QPC current data points.
This can be understood from simple qualitative arguments:
When $\gamma_f \Delta t \ll 1$, the two probability distributions corresponding to different initial SET states 
will not have yet separated, the QPC cannot resolve between the two SET states, and the measurement error is large.
If, on the contrary, the sampling time is so large that $\Gamma_\alpha \Delta t \gg 1$, the QPC will average over several SET jumps (telegraph noise~\cite{jordan2004a}).
Therefore, to use the QPC as a detector we have to obey the hierarchy that $\gamma_f \Delta t \gg 1$ while $\Gamma_\alpha \Delta t \ll 1$.
Finite $\Delta t$ however, will always imply a measurement error.
The Full Counting Statistics is a useful tool to classify these different error contributions quantitatively~\cite{schaller2010c}.

We can ask for the probability $P_F^{\rm err}$ of erroneously measuring a large current when the SET is filled throughout $\Delta t$
or for the probability $P_E^{\rm err}$ of erroneously measuring a small current when the SET is empty throughout.
The expected values for the currents would be $I_F = \kappa \gamma_f$ and $I_E = \gamma_f$, respectively, 
and it seems reasonable to put a discrimination threshold
right in the middle between these bounds, i.e., at
$m_{\rm th} = \gamma_f \Delta t(1+\kappa)/2$ (compare dashed line in Fig.~\ref{FIG:trajectory_example}).
The resulting error probabilities of maintaining the initial SET state throughout $\Delta t$ while measuring 
an erroneous QPC current can be calculated by using the counting fields $\chi$ and $\xi$ to
obtain the probabilities $P_{0m|E/F}(\Delta t)$ of zero SET jumps and $m$ QPC charge transfers during $\Delta t$ for an empty
or filled SET
\bea\label{EQ:det_errors}
P_E^{\rm err} &=& \sum_{m=0}^{m_{\rm th}} P_{0m|E}(\Delta t) = e^{-{\cal L}^{10}\Delta t} \frac{\Gamma(1+m_{\rm th}, \gamma_f \Delta t)}{\Gamma(1+m_{\rm th})}\,,\nn
P_F^{\rm err} &=& \sum_{m=m_{\rm th}+1}^\infty P_{0m|F}(\Delta t)\nn
&=& e^{-{\cal L}^{01}\Delta t}\left[1 - \frac{\Gamma(1+m_{\rm th}, \kappa \gamma_f \Delta t)}{\Gamma(1+m_{\rm th})}\right]\,.
\eea
Here, $\Gamma(x)$ denotes the $\Gamma$-function and $\Gamma(x,y)$ the incomplete $\Gamma$-function~\cite{arfken2005}.
These errors can be made small when $\gamma_f \Delta t > \kappa \gamma_f \Delta t\gg 1$, which can be achieved e.g. by increasing the QPC bias voltage, 
compare Eq.~(\ref{EQ:qpc_parameters}).

Furthermore, a good detector should also not miss any SET jump events. 
That is, we should allow for at most a single SET jump during $\Delta t$.
The corresponding error probability can be evaluated with the counting field $\chi$ and becomes
\bea
P_{\rm miss} &=& \sum_{n\ge2} \sum_m P_{nm}(\Delta t)\nn
&=& 1 - \frac{e^{-{\cal L}^{10} \Delta t} {\cal L}^{01} - e^{-{\cal L}^{01} \Delta t} {\cal L}^{10}}
{{\cal L}^{01}-{\cal L}^{10}}\,,
\eea
which happens to be independent of the initial SET occupation.
To make this term small, we require $\Gamma_\alpha \Delta t \ll 1$.

Combining these individual and mutually exclusive contributions, the total detection error becomes
\bea\label{EQ:detection_error}
P_{\rm err}(\Delta t) &=& P_E^{\rm err} P_E  + P_F^{\rm err} P_F + P_{\rm miss}\,,
\eea
where $P_E={\cal L}^{01}/({\cal L}^{01}+{\cal L}^{10})$ and $P_F=1-P_E$ are the (stationary) probabilities of
finding the SET empty or filled, respectively.
As all individual contributions are positive, they all constitute individual lower bounds to the detection error.

\section{Weak Measurements}\label{APP:measurements}

The counting statistics formalism is also useful to implement models for weak measurements.
To see that, it is useful to realize that the generalized master equation does not only propagate the system (in the 
main text implemented by the dot) but also describes the evolution of the number of transferred particles $n$.
The latter can be considered as the quantum number of a virtual detector device, such that at any time we may
describe the joint density matrix of system and detector by
\bea
\sigma(t) = \sum_{n,m} \rho^{(n,m)}(t) \otimes \ket{n}\bra{m}\,.
\eea
The reduced state of the system is obtained by tracing over the detector degrees of freedom
$\rho(t) = \sum_n \rho^{(n,n)}(t) \equiv \sum_n \rho^{(n)}(t)$ and thus only depends on the diagonal elements $n=m$.
These are precisely the ones that remain after a projective measurement of the detector quantum number, which upon
outcome $n$ yields
\bea\label{EQ:measurement_projective}
\sigma^{(n)} = \frac{\ket{n}\bra{n} \sigma \ket{n}\bra{n}}{\trace{\ket{n}\bra{n} \sigma}} = \frac{\rho^{(n)}}{\trace{\rho^{(n)}}} \otimes \ket{n}\bra{n}\,,
\eea
where the corresponding probability of obtaining measurement outcome $n$ is
given by $P_n = \trace{\rho^{(n)}(t)}$, compare also App.~\ref{APP:fcs}.
By tracing over the detector degree of freedom, we obtain the action of the measurement on the system only
\bea\label{EQ:povm_definition}
\rho^{(n)}(t) \hat{=} {\cal M}_n \rho(t)\,,
\eea
which we have written as a superoperator (calligraphic).
This is not necessarily projective.
It follows from properties of the discrete Fourier transform that $\sum_n {\cal M}_n = \f{1}$.
Directly after such a measurement, system and virtual detector are no longer entangled, and the detector
variable can be reset to zero (in the end, we are interested in measuring currents, i.e., changes of the particle
number versus time).
From the Full Counting Statistics we can conclude that for a system described by the generalized Liouvillian ${\cal L}(\chi)$, 
after an evolution of $\Delta t$, its conditional density matrix reads (right before the next measurement)
\bea\label{EQ:measurement_iteration}
\rho^{(n)}(t) &=& \int\limits_{-\pi}^{+\pi} e^{{\cal L}(\chi) \Delta t} e^{-\ii n \chi} \frac{d\chi}{2\pi} \rho(t-\Delta t)\nn
&=& \int\limits_{-\pi}^{+\pi} e^{{\cal L}(\chi) \Delta t} e^{-\ii n \chi} \frac{d\chi}{2\pi} e^{-{\cal L}(0) \Delta t} \rho(t)\,.
\eea
In the last step we have inserted the reverse propagator in ignorance of the detector evolution to separate the effect of the 
measurement from the evolution of the system.
Using Eq.~(\ref{EQ:measurement_iteration}) and~(\ref{EQ:povm_definition}) thus defines the measurement superoperators 
${\cal M}_n(\Delta t)$, which thereby depend implicitly on the measurement time $\Delta t$.

\section{Canonical vs. Grand Canonical ensembles}\label{APP:canonical}

In the treatment of the meso-reservoir state, we have put it in a grand-canonical equilibrium state throughout
\bea
\rho_{\rm isl}^{\rm gc} = \frac{e^{-\beta(H_{\rm isl} - \mu(t) N_{\rm isl})}}{Z_{\rm isl}}\,,
\eea
which was motivated by the assumption that both energy and particle numbers of the metallic island may in principle change.
For example, when $H_{\rm isl} = \sum_{k=1}^N \epsilon_k c_k^\dagger c_k$ and $N_{\rm isl} = \sum_k c_k^\dagger c_k$, the
condition that the value of the chemical potential $\mu(t)$ is fixed by the total particle number in the metallic island reads
\bea\label{EQ:chempot_determinant}
N &=& \sum_{k=1}^K \trace{c_k^\dagger c_k \rho_{\rm isl}^{\rm gc}} = \sum_{k=1}^K \frac{1}{e^{\beta(\epsilon_k-\mu_N)}+1}\nn
&=& \int \frac{{\cal D}(\omega)}{e^{\beta(\omega-\mu_N)}+1} d\omega\,.
\eea
Thereby, a discrete change in the particle number of the metallic island corresponds to a variation of its chemical potential as
discussed in Sec.~\ref{SEC:islandcharging}.
Similarly, a continuous change of the particle number can be mapped to a differential equation for the chemical potential as 
in Eq.~(\ref{EQ:diffeq_chempot}).

On the other hand, since the coupling between metallic island and its background reservoir is assumed significantly stronger than
the coupling between island and system, it would also be reasonable to treat -- between tunneling events via the island-system junction
-- the island in the canonical ensemble, i.e., 
\bea
\rho_{\rm isl}^{\rm cn} = \sum_{\f{n}:N_{\f{n}}=N} \frac{e^{-\beta E_{\f{n}}}}{Z_N}\,,
\eea
where with $\ket{\f{n}} \equiv \ket{n_1,n_2,\ldots,n_K}$ ($n_i \in \{0,1\}$) denoting all possible micro-states we have
$H_{\rm isl} \ket{\f{n}} = E_{\f{n}} \ket{\f{n}}$ with $E_{\f{n}} = \sum_{k=1}^K \epsilon_k n_k$ and $N_{\f{n}} = \sum_{k=1}^K n_k$.

\begin{figure}[ht]
\includegraphics[width=0.48\textwidth,clip=true]{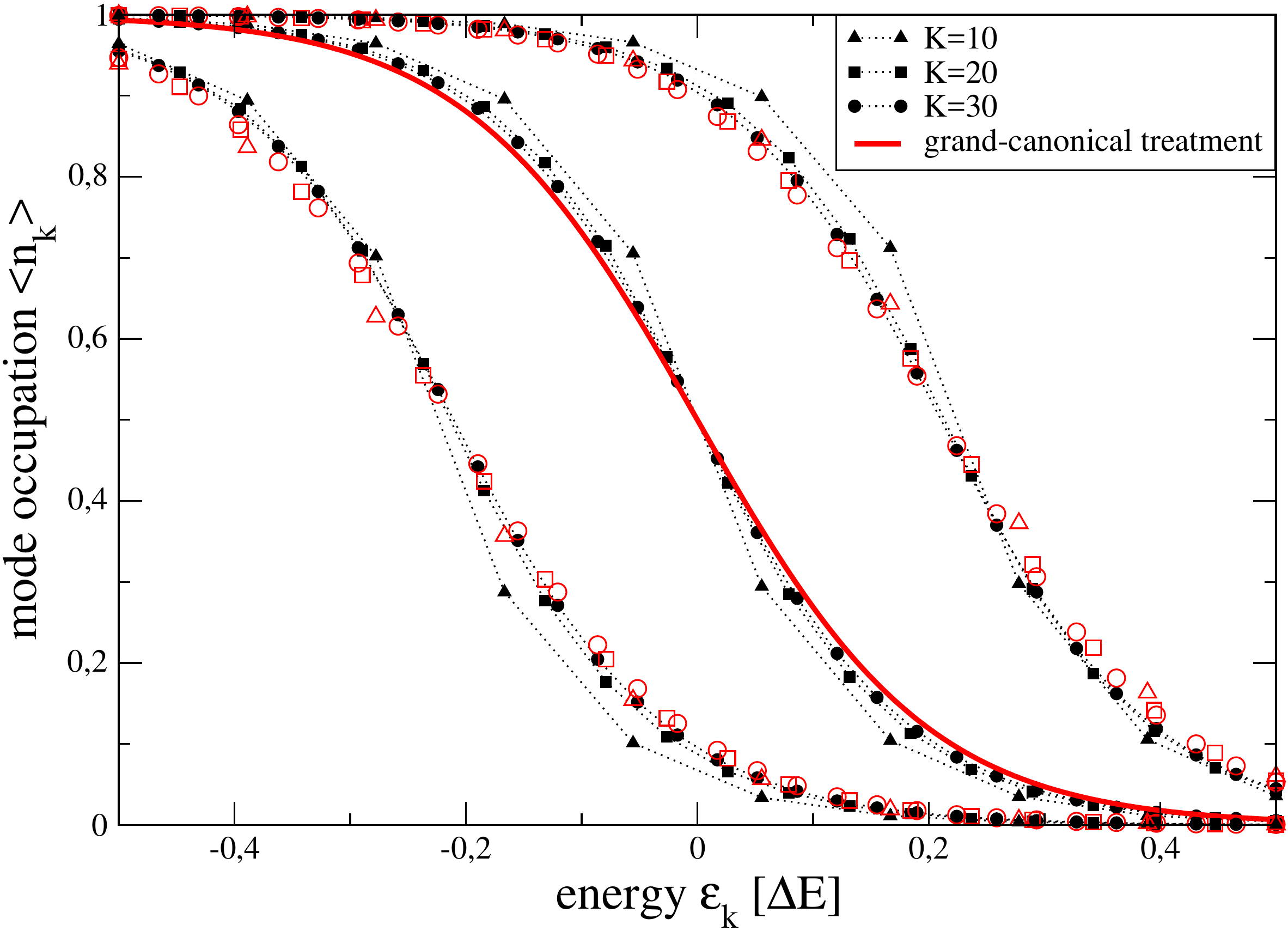}
\caption{\label{FIG:ensemble_comparison1}
Plot of the grand-canonical (hollow red symbols and red curve) and canonical occupations (solid black) versus energy 
levels for different mode numbers $K$ and 30\% filling ($N=0.3 K$, left), 50\% filling ($N=0.5 K$, middle), and 70 \% ($N=0.7K$, right).
Red and black circles are always much closer than e.g. red and black triangles, demonstrating convergence of canonical and grand-canonical 
treatments in the thermodynamic limit.
Other parameters: $\beta\Delta E=10$.
}
\end{figure}

Treating the metallic island in the grand-canonical and canonical ensemble is not fully equivalent. 
For example, the fluctuations of the total particle number $\expval{N^2}-\expval{N}^2$ are finite  in the grand-canonical treatment
but vanish in the canonical treatment.
By construction, the total particle number $\expval{N}$ is the same in both grand-canonical and canonical treatments.
The occupation of individual modes $\expval{c_k^\dagger c_k}$ may however be different.
To investigate this difference numerically, we choose for a given number of modes $K$ the distribution of
the single-particle energies as
\bea
\epsilon_k = \frac{\Delta E}{K-1} \left(k-\frac{K+1}{2}\right)\,,\qquad k\in\{1,\ldots,K\}\,,
\eea
such that we have $-\Delta E/2 \le \epsilon_k \le +\Delta E/2$.
This means that as $K$ increases, the splitting between levels becomes smaller and smaller, such that a continuum is reached as
$K\to\infty$. 
However, the spread of the levels remains constant at $\Delta E$.
We calculate $\expval{c_k^\dagger c_k}$ for all $k\in\{1,\ldots,K\}$ using both a grand canonical treatment by solving numerically Eq.~(\ref{EQ:chempot_determinant}) 
for $\mu_N$ (simple) and a canonical treatment $\expval{c_k^\dagger c_k} = \trace{c_k^\dagger c_k \rho_{\rm isl}^{\rm cn}}$
(computationally intensive).
We compare these distributions for different filling ratios and different mode numbers $K$, see Fig.~\ref{FIG:ensemble_comparison1}.
For half-filling, the grand-canonical occupation becomes extremely simple: Here, the chemical potential $\mu_{K/2}=0$ for all curves, 
and the grand-canonical occupation just becomes a simple Fermi function (solid red curve), sampled at the corresponding energies.
It is visible that already for moderate $K$, the difference between grand-canonical and canonical treatment is small, 
and the difference decreases with increasing number of modes.

Finally, we mention that scaling the total level spread $\Delta E \to \sqrt{K} \Delta E$, we obtain a model which is consistent with the 
wide-band limit used in the article.
This is equivalent with the results considered in Fig.~\ref{FIG:ensemble_comparison1} when the temperature is rescaled as $\beta \to \beta/\sqrt{K}$.

\section{Entropic balances of piecewise-constant rate equations}\label{APP:entropy_general}

We first briefly review the thermodynamics in absence of feedback actions and then turn toward its modification in presence of 
piecewise-constant feedback.

\subsection{In absence of feedback}

Numbering the system states by the index $i$, which are characterized by energies $E_i$ and particle numbers $N_i$, 
we consider systems governed by a rate equation of the form
$\dot{P}_i = \sum_\alpha \sum_j {W}_{ij}^{(\alpha)} P_j$.
Here, ${W}_{ij}^{(\alpha)} \ge 0$ is a transition rate from state $j$ to state $i$, that is triggered by a thermal reservoir $\alpha$.
The conservation of probabilities implies for independent reservoirs that $\sum_{i} {W}_{ij}^{(\alpha)}=0$ and hence we find that
${W}_{ii}^{(\alpha)} = -\sum_{j\neq i} {W}_{ji}^{(\alpha)}$.
Most important, for a thermal reservoir $\alpha$, the ratio of the associated backward and forward transition rates obeys in absence of feedback
detailed balance relations
\bea\label{EQ:detailed_balance_conventional}
\frac{W_{ij}^{(\alpha)}}{W_{ji}^{(\alpha)}} = e^{-\beta_\alpha [(E_i-E_j)-\mu_\alpha(N_i-N_j)]}\,.
\eea
When there is only a single reservoir, these relations imply that the grand-canonical equilibrium state $P_i \propto e^{-\beta (E_i - \mu N_i)}$ 
is a stationary solution of the rate equation.
For multiple reservoirs, these relations imply that the entropy production rate is positive
\bea\label{EQ:second_law_nofb}
\dot{S}_{\ii} = \dot{S} - \sum_\alpha \beta_\alpha \dot{Q}^{(\alpha)} \ge 0\,, 
\eea
which establishes the second law of thermodynamics.
Here, $S=-\sum_i P_i \ln P_i$ denotes the Shannon entropy of the system and $\dot{Q}^{(\alpha)}=\sum_{ij} [(E_i-E_j)-\mu_\alpha (N_i-N_j)] W_{ij}^{(\alpha)} P_j$ 
denotes the heat flow entering the system from reservoir $\alpha$.

\subsection{With feedback}

We now consider a feedback conditioned on the system being in state $j$.
Physically, this means that some external controller monitors the state of the system, and upon detecting the system in state $j$, it
immediately changes the system properties accordingly:
The energies of all levels $i$ are changed to $E_i^{(j)}$ and also the transition rates 
from $j$ to other states are changed to $W_{ij}^{(j,\alpha)}$.
Then, the rate equation under feedback becomes
\bea
\dot{P}_i = \sum_\alpha \sum_j {W}_{ij}^{(j,\alpha)} P_j\,.
\eea

As we will see, one can distinguish between changes of bare tunneling rates and changes of the energy levels.
Whereas the first type leaves the energetics of the system invariant but changes the entropy and
is for this reason also called Maxwell demon feedback~\cite{esposito2012a}, changing the energy levels modifies both the
energetic and entropic balances. 
It can therefore also not be considered a simple work source.

\begin{figure}[ht]
\begin{center}
\includegraphics[width=0.25\textwidth,clip=true]{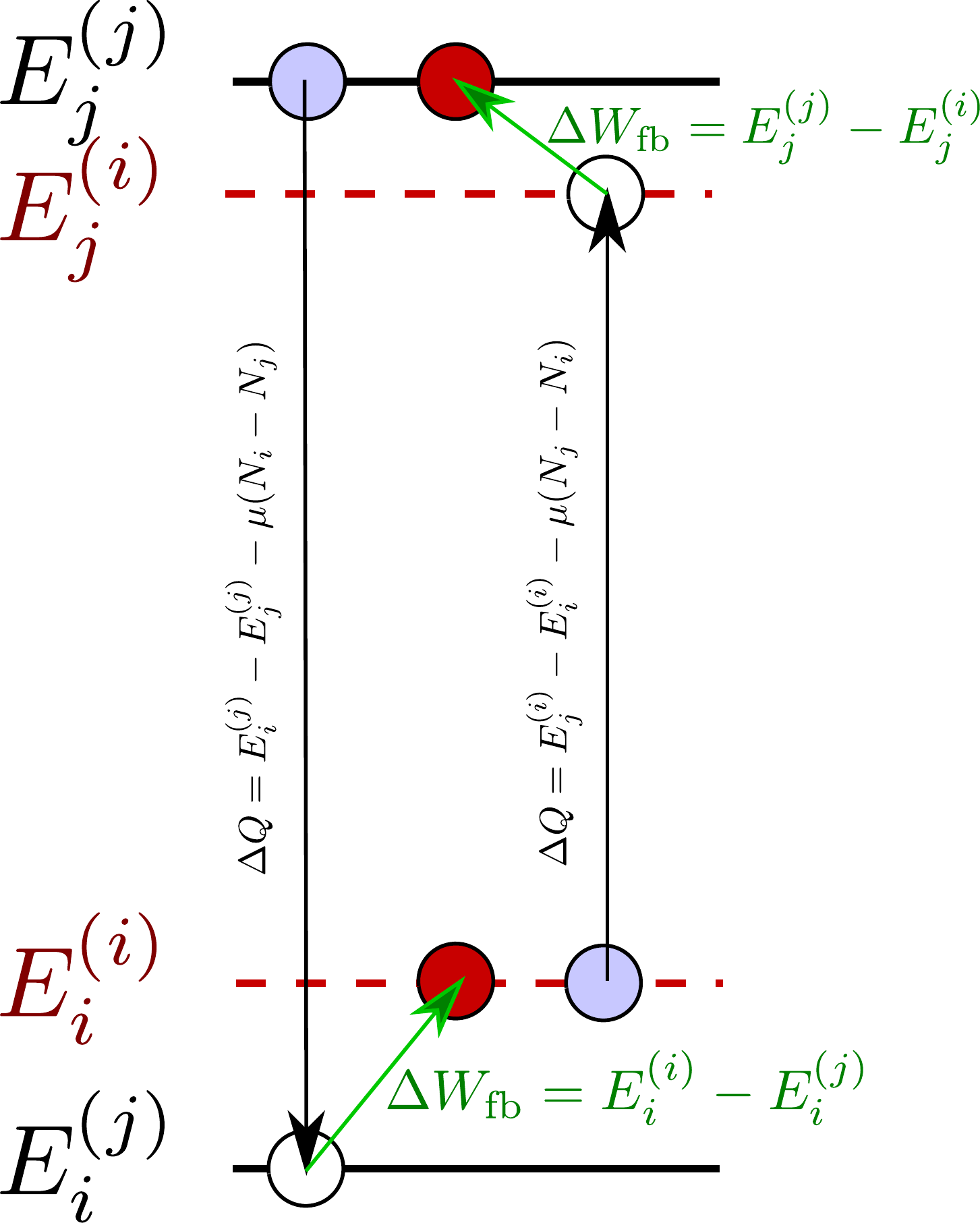}
\end{center}
\caption{\label{FIG:levelsketch}
Sketch of the energetic balance for the transition from from state $j\to i$ (left) and from state $i\to j$ (right) subject to feedback control
applied immediately thereafter.
The initial transition (blue to hollow circles) leads to the exchange of heat between system and reservoir (vertical terms).
Immediately thereafter, the control action changes the energy levels (hollow to filled red circes), thereby injecting energy into
the system if the level is occupied.
}
\end{figure}

During a jump $j\to i$ (where the system particle number changes according to $\Delta N_{ij} = N_i-N_j$), the energy balance of the system becomes 
$\Delta E_{ij} = (E_i^{(j)} - E_j^{(j)}) + (E_i^{(i)} - E_i^{(j)})$, where the first contribution is exchanged with the reservoir in form
of heat $\Delta Q_{ij} = (E_i^{(j)} - E_j^{(j)}) - \mu (N_i-N_j)$, and the second describes feedback energy $\Delta E_{\rm fb}$ injected into the system 
from the control action following immediately thereafter, see also Fig.~\ref{FIG:levelsketch} for an illustration.
This enables us to write the energy and particle currents entering the system from reservoir $\alpha$ as
\bea
I_E^{(\alpha)} &=& \sum_{ij} (E_i^{(j)}-E_j^{(j)}) W_{ij}^{(j,\alpha)} P_j\,,\nn
I_M^{(\alpha)} &=& \sum_{ij} (N_i-N_j) W_{ij}^{(j,\alpha)} P_j\,.
\eea
The energy injected in the system with the feedback actions can be similarly computed 
\bea
I_E^{\rm fb} = \sum_\alpha \sum_{ij} (E_i^{(i)}-E_i^{(j)}) W_{ij}^{(j,\alpha)} P_j\,,
\eea
and together we find for the total change of the system energy $E = \sum_i E_i^{(i)} P_i$
\bea\label{EQ:first_law}
\dot{E} &=& \sum_{ij} \sum_\alpha E_i^{(i)} W_{ij}^{(j,\alpha)} P_j\nn
&=& \sum_\alpha \sum_{i\neq j} E_i^{(i)} W_{ij}^{(j,\alpha)} P_j - \sum_\alpha \sum_{i\neq j} E_i^{(i)} W_{ji}^{(i,\alpha)} P_i\nn
&=& \sum_\alpha \sum_{i,j} (E_i^{(i)}-E_j^{(j)}) W_{ij}^{(j,\alpha)} P_j
= \left(\sum_\alpha I_E^{(\alpha)}\right) + I_E^{\rm fb}\nn
&=& \sum_\alpha \mu_\alpha I_M^{(\alpha)} + I_E^{\rm fb} + \sum_\alpha (I_E^{(\alpha)}-\mu_\alpha I_M^{(\alpha)})\,.
\eea
This is the first law of thermodynamics, where in the last line we can identify the chemical work done on the system, the energy injected from the feedback, and
the heat currents entering from the reservoirs.

We can also consider the evolution of the systems Shannon entropy $S=-\sum_i P_i \ln P_i$, where we get from algebraic manipulations~\cite{schaller2014}
\bea
\dot{S} &=& -\sum_i \dot{P}_i \ln P_i = \dot{S}_{\ii} + \dot{S}_{\rm e}\,,\nn
\dot{S}_{\ii} &=& \sum_\alpha \sum_{ij} W_{ij}^{(j,\alpha)} P_j \ln \left(\frac{W_{ij}^{(j,\alpha)} P_j}{W_{ji}^{(i,\alpha)} P_i}\right)\ge 0\,,\nn
\dot{S}_{\rm e} &= &\sum_\alpha \sum_{ij} W_{ij}^{(j,\alpha)} P_j \ln \left(\frac{W_{ji}^{(i,\alpha)}}{W_{ij}^{(j,\alpha)}}\right)\,.
\eea
Here, the positivity of the entropy production rate $\dot{S}_{\ii}$ follows from mathematical terms (it has the form of a relative entropy), 
and the second term $\dot{S}_{\rm e}$ can from the conventional detailed balance relation~(\ref{EQ:detailed_balance_conventional}) in absence of feedback 
be identified as the negative entropy change in the reservoirs.
However, the feedback changes the detailed balance relation in a way which we phenomenologically parametrize as
\bea\label{EQ:detailed_balance_feedback}
\frac{W_{ji}^{(i,\alpha)}}{W_{ij}^{(j,\alpha)}} = e^{\beta_\alpha [(E_i^{(j)}-E_j^{(j)})-\mu_\alpha(N_i-N_j)]} e^{-\Delta_{ij}^{(\alpha)}} e^{-\sigma_{ij}^{(\alpha)}}\,.
\eea
Here, the first term is associated with the entropy change of the reservoirs, indeed we can recover the heat flow from the reservoirs
into the system from it.
The second term $\Delta_{ij}^{(\alpha)}$ parametrizes changes of the transition rates that are not associated with energetic changes in the system.
Consequently, it must not depend on the reservoir temperatures.
Finally, the term $\sigma_{ij}^{(\alpha)}$ gathers all remaining influences of the feedback.
By distinguishing between $\Delta_{ij}^{(\alpha)}$ and $\sigma_{ij}^{(\alpha)}$ we have presupposed that an unambiguous discrimination between these
feedback effects is possible, see Sec.~\ref{SEC:entropic_balance} in the main text.
Inserting this decomposition into the ``entropy flow'' term we obtain
\bea
\dot{S}_{\rm e} &=& \sum_\alpha \beta_\alpha \dot{Q}^{(\alpha)} - {\cal I}_1-{\cal I}_2\,,\nn
{\cal I}_1 &=& \sum_\alpha \sum_{ij} W_{ij}^{(j,\alpha)} P_j \Delta_{ij}^{(\alpha)}\,,\nn
{\cal I}_2 &=& \sum_\alpha \sum_{ij} W_{ij}^{(j,\alpha)} P_j \sigma_{ij}^{(\alpha)}\,.
\eea
Solving for the entropy production, we can express it as
\bea\label{EQ:second_law}
\dot{S}_{\ii} = \dot{S} - \sum_\alpha \beta_\alpha \dot{Q}^{(\alpha)} + {\cal I}_1 + {\cal I}_2 \ge 0\,.
\eea
This is the second law of thermodynamics in presence of a non-equilibrium environment and feedback control.

At steady state, $\dot{S}\to 0$, and the usual inequality for the currents~(\ref{EQ:second_law_nofb}) is modified by two effective currents.
The first one ${\cal I}_1$ is associated with feedback actions that have no direct impact on the energetics, whereas the
second one takes the energetic feedback actions into account.
We note here that these information currents are just an effective description (for example, they can become negative), since we have not 
made the feedback loop explicit in our treatment but remain at a phenomenologic level.
If that is done for a microscopic treatment of the detector~\cite{strasberg2013a}, it is possible to link the effective information 
current with the time-derivative of the mutual information between controlled system and detector device~\cite{barato2013a,horowitz2014a}.

Depending on the regime, one may identify contributions to the total entropy production rate~(\ref{EQ:second_law}) which are negative.
These always need to be compensated by the other, positive contributions, which enables one to define information-theoretic efficiencies that
are upper-bounded by one.

\end{appendix}
\end{document}